\documentclass[preprint,groupedaddress,amsfonts,amssymb,amsmath]{revtex4-1}
\pdfoutput=1
\usepackage{bm}
\usepackage{graphicx}
\usepackage{tabularx}
\usepackage{textcomp}
\usepackage{threeparttable}
\usepackage{times}
\usepackage{units}
\usepackage{xspace}
\usepackage{setspace}
\usepackage{subfigure}

\providecommand{\T}[1]{\overleftrightarrow{\boldsymbol #1}}
\providecommand{\Av}[1]{\left\langle #1 \right\rangle}
\providecommand{\tx}[1]{\text{#1}}

\begin{document}

\title{How the diffusivity profile reduces the arbitrariness of protein folding free energies}

\author
{M. Hinczewski$^{1,2}$, Y. von Hansen$^{1}$, J. Dzubiella$^{1}$, R. R. Netz$^{1\ast}$\\
\normalsize{$^{1}$Physics Department, Technical University of Munich, 85748 Garching, Germany}\\
\normalsize{$^{2}$Institute for Physical Science and Technology, University of Maryland, College Park,
MD 20742}\\
\normalsize{$^\ast$To whom correspondence should be addressed; E-mail:  netz@ph.tum.de}\\}

\begin{abstract}

The concept of a protein diffusing in its free energy folding
landscape has been fruitful for both theory and experiment. Yet the
choice of the reaction coordinate (RC) introduces an undesirable
degree of arbitrariness into the problem.  We analyze extensive
simulation data of an $\alpha$-helix in explicit water solvent as it
stochastically folds and unfolds.  The free energy profiles for
different RCs exhibit significant variation, some having an activation
barrier, others not. We show that this variation has little effect on
the predicted folding kinetics if the diffusivity profiles are
properly taken into account.  This kinetic quasi-universality is
rationalized by an RC rescaling, which, due to the reparameterization
invariance of the Fokker-Planck equation, allows the combination of
free energy and diffusivity effects into a single function, the
rescaled free energy profile.  This rescaled free energy indeed shows
less variation among different RCs than the bare free energy and
diffusivity profiles separately do, if we properly distinguish between
RCs that contain knowledge of the native state and those that are
purely geometric in nature.  Our method for extracting diffusivity
profiles is easily applied to experimental single molecule time series data and might help to
reconcile conflicts that arise when comparing results from different experimental
probes for the same protein.
\end{abstract}
\maketitle
\section{Introduction}

The problem of protein folding kinetics is formidable from a purely
statistical mechanics point of view: The unfolded protein, in other
words the entire ensemble of micro-states that significantly deviate
from the native state, transits via a myriad of distinct pathways to
the folded (native) state, and trying to predict folding times from
basic principles is obviously hopeless.  Yet, robust features have
emerged both from experiments and theoretical
concepts~\cite{review1,review2}.  A key fact is that any experiment
that probes protein folding or unfolding projects protein micro-states
onto a low-dimensional (typically one-dimensional) observable.  For
example, circular dichroism in the far ultraviolet and infrared
adsorption spectroscopy basically measure the average helicity, while
fluorescence is sensitive to side chain contacts or local solvent
structure around tryptophan residues~\cite{Wolynes,Fersht1}.  Kinetic
information at ambient conditions and on short time scales relevant
for fast folding events can be obtained by time-resolved spectroscopy
after flash photoheating~\cite{Eaton3} or by FRET and TTET correlation
studies that couple to the distance between a donor and acceptor
linked to two positions along the peptide chain~\cite{Kief1,Kief2}.
More recently, single-molecule spectroscopic techniques have allowed
the observation of time-dependent folding/unfolding of individual
proteins, thus going beyond ensemble averaging~\cite{Schuler,Eaton2}.
Likewise, single molecule studies where forces are applied at two
points along the peptide backbone probe the distance between those two
anchoring points~\cite{Rief1}.  All these experimental observables in
fact constitute distinct {\em reaction coordinates} (RCs).

Exponential distributions of folding times found for many (but not
all) proteins using different techniques suggest two-state-folding as
a quite general paradigm of folding kinetics: here the folded and
unfolded states are separated by a free energy barrier along the
respective RC~\cite{Fersht1}. 
Even proteins folding via many intermediate states
can produce a single exponential folding time if there
exists a rate-limiting transition. Therefore,
as long as the reaction coordinate of choice distinguishes the two
states connected by the rate-limiting step, using different kinds of
measurement/reaction coordinate would likely generate similar
single-exponential kinetics even in such a case.
Similar conclusions can been drawn from
the direct observation of population distributions, where a free
energy barrier means that folding intermediates are rarely
observed~\cite{Schuler,Eaton2}.  The recent observation that different
experimental techniques yield different kinetics~\cite{Gruebele} or
distribution functions~\cite{Munoz,Fersht2} when applied to the same
protein casts doubt on the clear division between two-state
(exhibiting a free-energy barrier) and down-hill folders (without such
a barrier).  In this paper we argue that such inconsistencies can
arise when implicitly referring to different RCs, and show a way of
how to reconcile conflicting results.
 
In theoretical studies, various RCs have become popular to
characterize the folding transition, either because they approximately
correspond to an experimentally accessible observable or because they
are simple to calculate.  The radius of gyration, the fraction of
native contacts between residues, or the mean distance from the native
state are typical examples~\cite{Thirumalai1,Thirumalai2}.  More
complex topological order parameters such as the contact order have
been suggested for describing universal features of protein folding
kinetics~\cite{Plaxco}.  In the theoretical framework that naturally
emerges, the protein diffuses along the RC, governed by a stochastic
equation and subject to deterministic forces encripted in the free
energy landscape, as well as stochastic forces due to the random
environment~\cite{Hopfield,Socci,Grosberg}.  Early on, it was realized
that the diffusion constant in this coarse-grained picture is an
effective quantity that takes into account the connectivity between
states (i.e. the number of possible connecting paths), the energetic
ruggedness of such paths~\cite{Zwanzig1}, as well as orthogonal
degrees of freedom~\cite{Zwanzig2}.  As folding progresses, internal
friction starts to play a more dominating role~\cite{Weeks,Alfredo},
while solvent friction becomes less important as more and more peptide
groups lose solvent contact~\cite{Eaton3}.  Recently, the
simplification of a constant diffusivity was abandoned and a
diffusivity profile was extracted from simulations of peptides: these
works either considered proteins without solvent (and thus exclude
variations of the solvent friction)~\cite{BestHummer,Wang,BestHummer3}
or considered exclusively short-time dynamics and thus are not
applicable to global folding kinetics~\cite{Onuchic}.  The trifold
coupling between the choice of a specific RC and the free energy and
diffusivity profiles in the presence of explicit solvent has remained
elusive.

In this paper we perform an in-depth analysis of long MD trajectories
of an $\alpha$-helix forming oligo-peptide including explicit
water.  Such model peptides form the subject of detailed experimental
studies and constitute some of the simplest peptides that exhibit
non-trivial folding kinetics~\cite{Baldwin}.  They are thus
interesting in their own right and at the same time---due to their
minute size---allow for realistic modelling over times much longer than
their folding times, including solvent degrees of freedom~\cite{Joe}.
As a prerequisite for our analysis, we introduce a simple way of
extracting diffusivity profiles from time series data for an arbitrary
RC, that can be conveniently applied to experimental spectroscopic
data~\cite{Eaton2}, or force spectroscopic data for RNA~\cite{Block},
or proteins~\cite{Rief3} as well.
We demonstrate that different RCs for one and the same protein
trajectory are associated with substantially different free energy
profiles, some showing a barrier separating the folded and unfolded
helix state, some showing no barrier at all (which is not surprising
and has been found in different contexts before~\cite{Thirumalai3}).
This resembles the experimental findings in connection with the
dispute on down-hill versus two-state folding~\cite{Munoz,Fersht2},
but is resolved by accounting for the spatially inhomogeneous
diffusivity: The diffusivity profiles are full of structure and show
considerable variation among different RCs.  No simple connection
between the  free energy and diffusivity profiles seems to exist.  
Yet, the folding kinetics predicted using a stochastic approach based on the
free energy landscape is largely independent of the RC if and only if
the diffusivity profile is taken into account.  Thus, the variance
between free energy profiles along different RCs gives rise to kinetic
universality if the coupling to diffusivity is included (where we
distinguish between reaction coordinates that contain knowledge of the
native state and those that are purely geometric in nature).  This
specifically means that the presence of a free energy barrier
(i.e. absence of intermediate states) is in principle compatible with
both exponential and non-exponential kinetics, and that different
experimental probes are bound to measure different free energy
profiles. The same conclusions also apply to more refined or optimized
RCs~\cite{Bolhuis,BestHummer2,Eric,Orland,Noe}.  Full understanding of protein
folding kinetics thus requires measuring both average distributions
and kinetic trajectories.  Similar conclusions were very recently
drawn from a Bayesian analysis of folding trajectories of simple
coarse-grained model peptides based on implicit-solvent
simulations~\cite{BestHummer3}.  Since $\alpha$-helices are a
prominent folding motif, the features we find are most likely relevant
for more complex proteins as well.

\section{Methods}

\noindent {\bf Simulations -} Standard all-atom MD simulations provide 1.1
$\mu$s trajectories of an alanine (A)-based peptide with sequence
Ace-AEAAAKEAAAKA-Nme in explicit water~\cite{Joe}, which is a
shortened version of similar sequences with charged Glu$^+$ (E) and
Lys$^-$ (K) residues at positions $i$ and $i+4$ that experimentally
are known to spontaneously form $\alpha$-helices~\cite{Baldwin}.  The
mechanism for $\alpha$-helix formation involves, in addition to the
stabilizing influence of E-K salt bridges, hydration
effects~\cite{Ghosh,Joe}.  The MD simulations utilize the parallel
module sander.MPI in the Amber 9.0 package with the ff03 force-field
and the TIP3P water model at a pressure of 1 bar and a temperature $T$
fixed by a Berendsen barostat and Langevin thermostat,
respectively~\cite{amber}.  The periodically repeated cubic simulation
box has an edge length $L \approx 36$\AA\ including $\approx$ 1500
water molecules.  Electrostatic interactions are calculated by
particle mesh Ewald summation and real-space electrostatic and van der
Waals interactions are cut off at 9 \AA.  As a check on the
convergence of the standard MD simulation, replica-exchange MD (REMD)
simulations are performed with the AMBER10 simulation
package~\cite{amber}.  Here the same force-field and system parameters as
in the other standard MD simulations are employed, apart from switching to a
constant volume ensemble.  32 replicas are considered in a temperature
range between 265 and 520 K, with each replica simulated for 22.5 ns,
amounting to a total sampling time of 720 ns. Temperature exchanges
between neighboring replicas are attempted every 250 integration
steps, leading to an exchange rate of 10 - 30\%.  \\[0.5em]
%
%
\begin{table}[t!]
\centering
\caption{List of reaction coordinates (RCs) used in the paper.} 
\begin{center}
\begin{tabular}{cc}
\hline
\hline
 RC notation  & description \\    
\hline
$Q_1$ &  RMS deviation from perfect helix \\
$Q_2$ &  native intra-backbone hydrogen bond length   \\
$Q_3$ &   inverse native hydrogen bond length    \\
$Q_4$ &   radius of gyration  \\
$Q_5$ &    end-to-end distance \\
\hline
\hline									
\end{tabular}
\end{center}
\label{RCdefinition}
\end{table}%
%
%
{\bf Reaction coordinates -} Trajectory analysis is
performed using the {\it ptraj} tool in the Amber
package.~~\cite{amber} The helicity (i.e., the $\alpha$-helical
fraction) is identified using the DSSP method by Kabsch and
Sander~~\cite{dssp}.  In addition, we focus on five different RCs to
follow the folding kinetics:

 (i) $Q_1$, defined as the
root-mean-square distance from a fully helical reference structure,
averaged over all $M$ atoms of the peptide.  The reference structure
was chosen randomly from configurations which display 100\% helicity,
with little variation depending on the specific choice.  

(ii) The mean
native hydrogen bond (HB) length, $Q_2 = \sum_{i=1}^{N-4}
r_{i,i+4}/(N-4)$, averaged over all $N=14$ residues including the
acetyl (Ace) and amine (Nme) end caps, where $r_{i,j}$ is the distance
between HB forming atoms, $i$ and $j$, in the peptide backbone.  

(iii) The mean inverse HB length, $Q_3 =1-(N-4)^{-1}\sum_{i=1}^{N-4}
r^0_{i,i+4}/r_{i,i+4}$, where $r^0_{i,i+4} \approx 2$ \AA\ is the
native HB length in the folded state, defined by the most probable
length of each $(i,i+4)$ HB.  

iv) The radius of gyration, $Q_4 =
\left[ \sum_{i,j=1}^M r_{i,j}^2 /(2M^2) \right]^{1/2}$, a measure for
the average peptide size and accessible in scattering.  

v) $Q_5$, the
distance between the centres of mass of the end caps.  Trajectories
are recorded with a resolution of 20~ps, giving a total of 54171 data
points.  To compare different RCs with each other, we exclude for each
RC the 11 smallest and 11 largest values, and define rescaled RCs 
\begin{equation}
q_i = (Q_i - Q_i^\text{min})/(Q_i^\text{max}-Q_i^\text{min})
\end{equation}
 such that
the minimal and maximal values of the remaining 54149 data points,
denoted as $Q_i^\text{min}$ and $Q_i^\text{max}$, are projected on the
RC values $q_i=0$ and $q_i=1$, respectively.\\[0.5em]
\begin{figure*}[t]
\centering
\includegraphics*[width = 15cm]{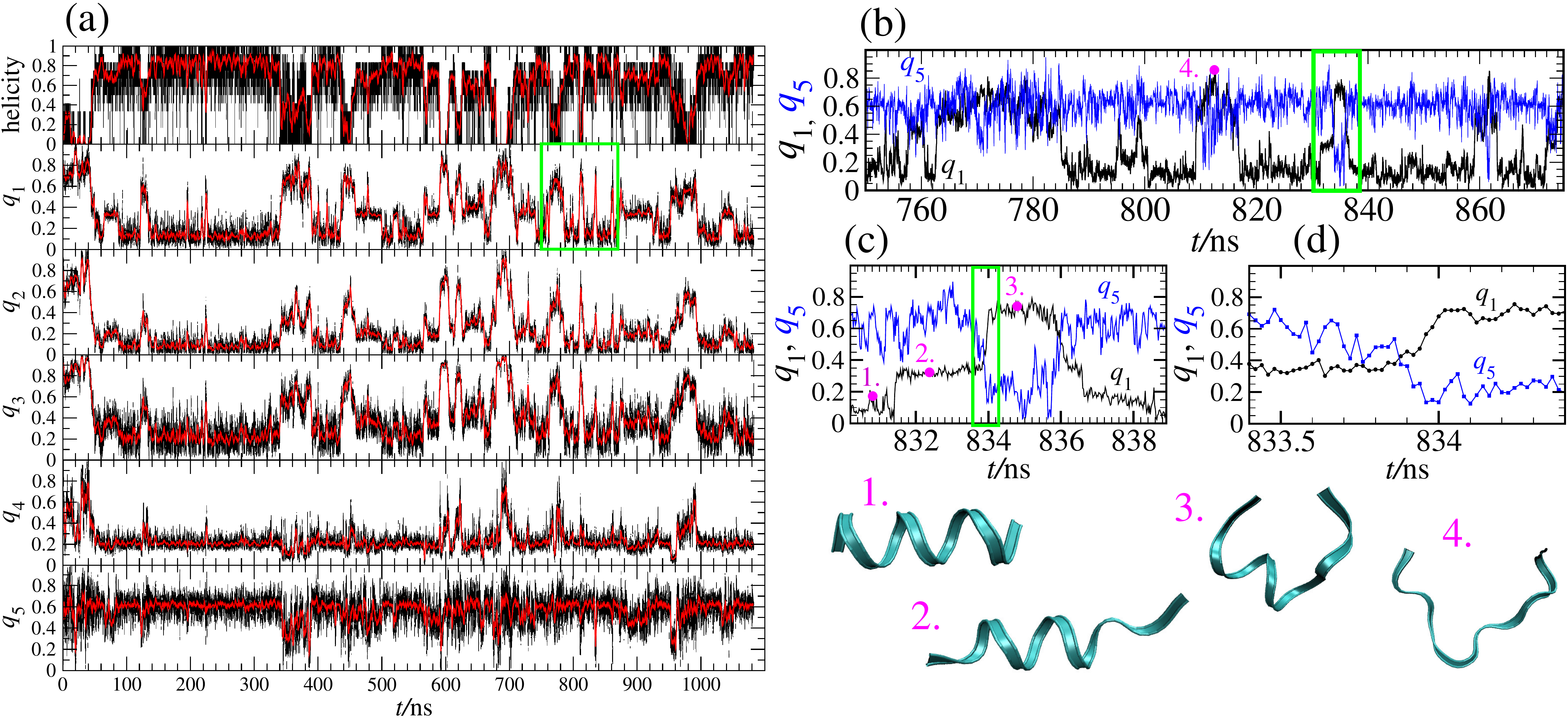}
\caption{Complete time series data of the simulation run for the peptide in explicit water.
Shown are helicity and the five considered RCs defined in Table I. Lines in black/blue  show the full resolution 
data (20 ps), while red lines are smoothed over time  windows of 2 ns. The right panels show selected data 
windows at higher time resolution for $q_1$ and $q_5$  together  with a few selected MD snapshots of 
the peptide backbone structure. }
\label{fig1}
\end{figure*}
{\bf Diffusion constant -} We assume that the stochastic
time evolution of a given RC is described by the one-dimensional
Fokker-Planck (FP) equation~\cite{Haenggi}
\begin{equation}\label{eq:1}
\frac{\partial}{\partial t} \Psi(Q,t) = \frac{\partial}{\partial Q}
 D(Q)   e^{-\beta F(Q)}  \frac{\partial}{\partial Q}  \Psi(Q,t) e^{\beta F(Q)}  
\end{equation}
where $\Psi(Q,t)$ is the probability of having a
configuration with RC value $Q$ at time $t$, $D(Q)$
is the (in general $Q$-dependent) diffusivity,
$\beta=1/(k_BT)$ and $\beta F(Q) = - \ln \langle \Psi(Q) \rangle $ is the free energy
where $ \langle  \Psi(Q) \rangle $ is the time-averaged 
probability distribution.
A few methods to extract $D(Q)$ from time-series data
based on Bayesian analysis of transition rates~\cite{Hummer,BestHummer2}
or short-time fluctuations have been described~\cite{Onuchic,Wang}.
Our method extracts $D(Q)$ directly from folding times.
Define $\tau_\text{FP}(Q, Q^f)$ as
the mean first passage (MFP) time to go from a state $Q$ to some
final  state $Q^\text{f}$ without recrossing $Q^\text{f}$,
corresponding to  an adsorbing boundary condition at $Q^\text{f}$.
For the case $Q>Q^f$ one finds~\cite{Weiss}
\begin{equation}\label{eq:2}
\tau_\text{FP}(Q, Q^f) = \int_{Q^\text{f}}^{Q} dQ^\prime\,\frac{e^{\beta F(Q^\prime)}}{D(Q^\prime)} 
\int_{Q^\prime}^{Q^\text{max}} dQ^{\prime\prime} e^{-\beta F(Q^{\prime\prime})}
\end{equation}
and for $Q<Q^f$ one has
\begin{equation}\label{eq:2b}
\tau_\text{FP}(Q, Q^f) = \int^{Q_\text{f}}_{Q} dQ^\prime\,\frac{e^{\beta F(Q^\prime)}}{D(Q^\prime)} 
\int^{Q^\prime}_{Q^\text{min}} dQ^{\prime\prime} e^{-\beta F(Q^{\prime\prime})},
\end{equation}
where  at $Q^\text{min}$ and $Q^\text{max}$ reflective (zero-flux) boundary conditions hold.
 By differentiation with respect to $Q$, we obtain the diffusivity for $Q>Q^f$ 
\begin{equation}\label{eq:3}
D(Q) = \frac{e^{\beta F(Q)}}{\partial \tau_\text{FP}(Q, Q^f) / \partial Q}
\int_{Q}^{Q^\text{max}} dQ^{\prime} e^{-\beta F(Q^{\prime})}
\end{equation}
and for $Q<Q^f$ as
\begin{equation}\label{eq:3b}
D(Q) = -  \frac{e^{\beta F(Q)}}{\partial \tau_\text{FP}(Q, Q^f) / \partial Q}
\int^{Q}_{Q^\text{min}} dQ^{\prime} e^{-\beta F(Q^{\prime})}.
\end{equation}
An even simpler procedure employs the {\em round-trip time}
\begin{equation}
\tau_\text{RT}(Q, Q^f) =\text{sign}(Q-Q^f) [ \tau_\text{FP}(Q, Q^f) + \tau_\text{FP}(Q^f, Q)],
\end{equation}
the magnitude of which is 
the time needed to start at $Q$,  reach $Q^f$ for the first time, start from $Q_f$ again
and reach back to $Q$ for the first time. One finds
\begin{equation}\label{eq:4}
\tau_\text{RT}(Q, Q^f) = Z  \int_{Q^\text{f}}^{Q} dQ^\prime\,\frac{e^{\beta F(Q^\prime)}}{D(Q^\prime)} 
\end{equation}
where $Z =\int_{Q^\text{min}}^{Q^\text{max}} \text{d} Q e^{-\beta F(Q)}$ is
the partition function.
The diffusivity profile based on the round-trip time reads
\begin{equation}\label{eq:5}
D(Q) = \frac{Z e^{\beta F(Q)}}{\partial \tau_\text{RT}(Q, Q^f) / \partial Q}.
\end{equation}
Intuitively, the slope of the round-trip time function is inversely
proportional to $D(Q)$: For a given $F(Q)$, a larger slope implies a
slower return to the starting point, or equivalently a smaller local
diffusivity.  The FP approach assumes an underlying Markovian process,
meaning that $D(Q)$ and thus $\partial \tau_\text{RT}(Q, Q^f) /
\partial Q $ are independent of $Q^f$.  We exploit (and check) this by
defining a mean round-trip time function $\bar{\tau}_\text{RT}(Q)$
that results from an average of round-trip times $\tau_\text{RT}(Q,
Q^f)$ over their final states $Q^f$.  Since on the FP level $\tau_\text{RT}(Q, Q^f)$
curves for different $Q^f$ differ only by an additive constant, we
should be able to collapse all such curves onto
$\bar{\tau}_\text{RT}(Q)$.  The assumption of Markovian behavior
breaks down at short times and for unsuitable reaction coordinates
(i.e. RCs that do not single out the transition state, as will be
explained in detail later on)
and is clearly indicated  by deviations of the  round-trip time functions
for varying $Q^f$,
$\tau_\text{RT}(Q, Q^f)$, from the mean  $\bar{\tau}_\text{RT}(Q)$.
  Insight into this can be gained with a simpler
definition of the diffusivity based on the variance in RC
space~\cite{Onuchic}
\begin{equation}\label{eq:6}
D_\text{var}(Q_0, \delta t) = \langle (Q(\delta t,Q_0) - \langle Q(\delta t,Q_0) \rangle )^2 \rangle /(2 \delta  t)
\end{equation}
where $Q(\delta t,Q_0)$ denotes one specific realization of a path
that starts at $Q_0$ at time $\delta t=0$.  As we will demonstrate,
$D_\text{var}(Q_0, \delta t)$ sensitively depends on the lag time
$\delta t$.  To get accurate results, $\delta t$ should be small
enough that the region explored by the RC in this time interval has an
approximately constant free energy; however if $\delta t$ is below a
threshold time scale, the resulting $D_\text{var}$ may be dominated by
non-Markovian properties.  We will mostly use the round-trip method
for determining $D(Q)$, but compare to the other methods as well.

In our analysis of the simulation time series data 
we discretize RCs in typically $K=50$ intervals
and normalize probability distributions 
according to  $\sum_{k=1}^{K} \Psi(Q^{(k)},t)=K$.\\[0.5em]
{\bf Fit of round-trip times -} To extract $D(Q)$ from the
simulation data requires estimating the derivative $\partial
\bar{\tau}_\text{RT}(Q) / \partial Q $.  We start by fitting a smooth
function to the numerical results, exploiting the fact that
$\bar{\tau}_\text{RT}(Q)$ should be a monotonically increasing
function of $Q$.  Thus the fitting function
$\bar{\tau}_\text{RT,fit}(Q)$ can be expressed in the form:
\begin{equation}
\bar{\tau}_\text{RT,fit}(Q)  = \bar{\tau}_\text{RT,fit}(Q^\text{min})  + \int_{Q^\text{min}}^{Q} dQ^\prime \,e^{W(Q^\prime)},
\end{equation}
where $W(Q^\prime)$ is an arbitrary function.  We expand out
$W(Q^\prime)$ in a basis of cubic B-splines defined over the range
$Q^\text{min}$ to $Q^\text{max}$, and use the coefficients of the
expansion as fitting parameters.  The size of the basis is fixed at
$40$ splines.  The full expression for $\bar{\tau}_\text{RT,fit}(Q)$
is fit to the simulation estimate for $\bar{\tau}_\text{RT}(Q)$ using
a standard least squares technique, with one modification: the
quantity to be minimized is the sum of squared residuals plus another
term which penalizes roughness in the fitted function.  This
additional term has the form $\lambda
\int_{Q^\text{min}}^{Q^\text{max}} dQ^\prime\,(\partial
W(Q^\prime)/\partial Q^\prime)^2$, with smoothing parameter $\lambda$.
Larger values of $\lambda$ lead to progressively smoother fits to the
data.  The entire fitting procedure is implemented through the
Functional Data Analysis package in the R programming
language~\cite{FDA}.  For all the results shown below we set $\lambda
= 50$, since we found that varying $\lambda$ in the range 10-200 had
minimal effect on the resulting diffusion profiles.  The range
$\lambda \ll 10$ is unsuitable because we fit to jagged features in
the simulation $\bar{\tau}_\text{RT}(Q)$ curve which are the result of
statistical noise.  For the range $\lambda \gg 200$, we over-smooth
the curve, losing most of the local slope information and resulting in
poor fits to the round-trip function.\\[0.5em]
{\bf Reparameterization -} As is
well-known~\cite{Pande,Karplus}, the FP Eq.~(\ref{eq:1}) is invariant
under an arbitrary RC rescaling according to $\tilde{Q} = \tilde{Q}
(Q)$ if the functions $\Psi$, $F$, $D$ are simultaneously rescaled as
$\tilde{\Psi} = \Psi / \tilde{Q}'$, $ \tilde{F} = F + \beta^{-1} \ln
\tilde{Q}'$, and $\tilde{D} = (\tilde{Q}')^2 D$. Here, $\tilde{Q}' =
\text{d} \tilde{Q}(Q) / \text{d}Q$ is assumed positive.  Thus an
arbitrary diffusivity profile $\tilde{D}(\tilde{Q})$ can be obtained,
while the kinetics on the FP level and the partition function $Z$ stay
invariant, as long as the folding free energy is adjusted
accordingly. For the particular choice of a constant diffusivity,
$\tilde{D} = \tilde{D}_0$, we get $\tilde{Q}' = \sqrt{\tilde{D}_0 /
  D}$ and thus $ \tilde{F} = F - ( 2 \beta)^{-1} \ln (D/\tilde{D}_0)$.
%
\begin{figure}[t!]
\centering
\includegraphics*[width = 8.2cm]{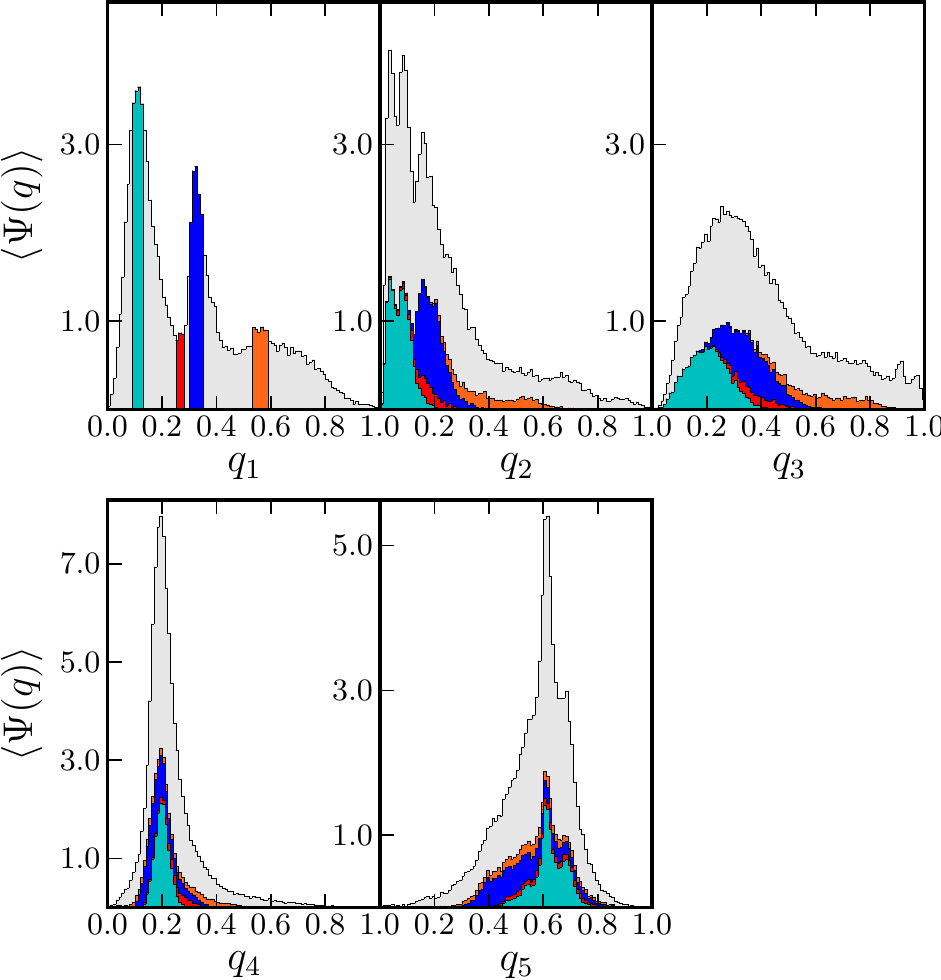}
\caption{ Mapping from RC $q_1$ to  different RCs.  Plotted is
  the mean distribution $\langle \Psi (q) \rangle$ for the entire time
  series data in Fig.~1 and---in different colours---selected regions
  of the distribution.  }
\label{fig2}
\end{figure}
\section{Results}

Fig.~1 shows the complete times series data for the simulated
oligopeptide. In all five RCs and in the helicity data frequent
switching between the folded state (large helicity and small $q_i$
values) and the unfolded state is observed, meaning that the
simulation is converged and allows drawing conclusions on the folding
and unfolding kinetics (further evidence is provided by the excellent
comparison between straight MD and replica-exchange simulations, as
shown in Fig.~6). The fine resolution data (Fig.~1, right panel) in
terms of the RMS-deviation from the fully helical state, RC $q_1$,
suggest that an intermediate state and two barriers are present.  As
the snapshots indicate, in the fully helical state ($q_1 \approx 0.1$)
roughly three $\alpha$-helical turns are stabilized by salt bridges
between the Glu$^+$-2 and Lys$^-$-6 and the Glu$^+$-7 and Lys$^-$-11
residues, respectively.  In the intermediate state ($q_1 \approx 0.4$)
only one of the two salt bridges stabilizes two turns, while in the
unfolded state ($q_1 \gtrsim 0.7$) no bridge is present.  Note that
the characteristic transition time for unfolding of one helical turn,
i.e. for the transition from $q_1\approx 0.4$ to $q_1 \approx 0.7$ in
(d), is roughly 200~ps and thus about 100 times shorter than the
corresponding unfolding time in Fig.~3(e).  While a high degree of
correlation between different RCs can be inferred from Fig.~1, there is
no one-to-one mapping, e.g., $q_5$ in Fig.~1(c) shows pronounced
fluctuations in intervals where $q_1$ stays virtually constant.

This is already evident from the average distribution function
$\langle \Psi (q) \rangle$ shown in Fig.~2 as a function of all
different RCs. While the distribution $\langle \Psi (q_1) \rangle$ in
the leftmost panel as a function of $q_1$ shows three broad peaks
(corresponding roughly to none, one and two intact salt bridges),
clearly separated peaks are absent when $\langle \Psi \rangle$ is
shown as a function of $q_2$, $q_3$, $q_4$ or $q_5$.  
The reason is simple: states that are
separated when, e.g., described by $q_1$, are mixed when they are
projected onto different RCs. This is demonstrated by the coloured
regions in Fig.~2 that for $q_1$ correspond to pure states, i.e.
narrow intervals of $q_1$ values. While for $q_2$ and 
$q_3$ the colored regions are smeared out but the
ordering along the RC is preserved, for $q_4$ and $q_5$ the ordering
is lost.  This points to a fundamental difference between the RCs
$q_1, q_2, q_3$, that embody knowledge of the native state, and the
RCs $q_4, q_5$, which are purely geometric.

\begin{figure}[t!]
\centering
\includegraphics*[width = 7.2cm]{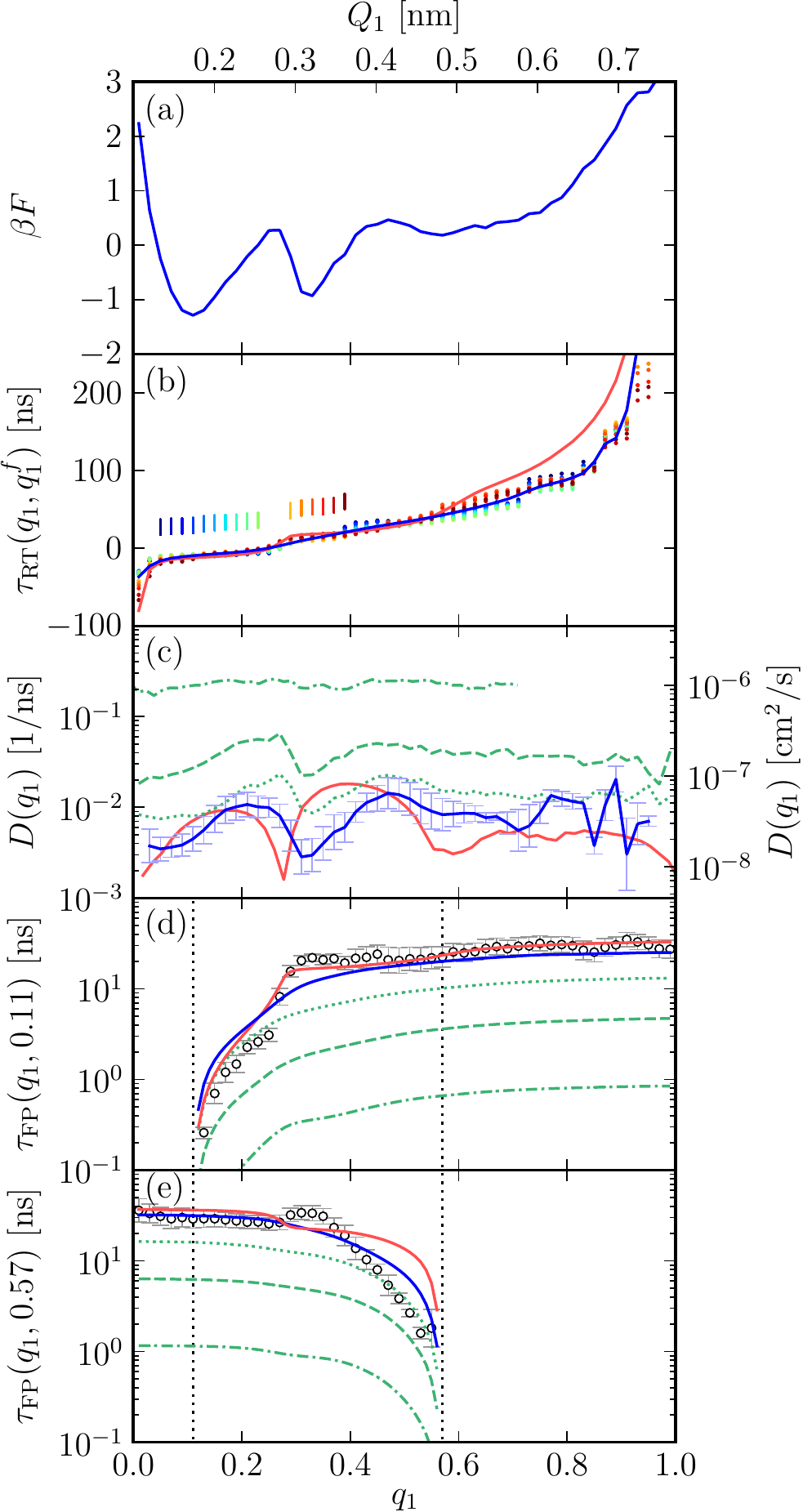}
\caption{ Results for RC $q_1$ (note that the upper scale is in terms
  of the unrescaled RC $Q_1$).  (a) Free energy profile $\beta F = -
  \ln \langle \Psi \rangle$.  (b) Data points give the round-trip
  times $\tau_\text{RT}(q_1, q_1^f)$ as extracted from the simulation
  data for various final states $q_1^f$ that are denoted by vertical
  colored bars. The data is shifted vertically for each $q_1^f$ to
  illustrate the theoretically predicted collapse onto a single mean
  round-trip curve $\bar{\tau}_\text{RT}(q_1)$, with the smooth fit
  $\bar{\tau}_\text{RT,fit}(q_1)$ shown in blue.  The red curve
  denotes the round-trip time from the Bayesian approach.  (c)
  Diffusivity from the round-trip time method Eq.~(\ref{eq:5}) (blue
  curve), compared to the variance method Eq.~(\ref{eq:6}) for lag
  times $\delta t=$200~fs, 20~ps, and 200~ps (dash-dotted, dashed,
  dotted green curves), and to the Bayesian method (red
  curve)~~\cite{Hummer}.  (d) MFP or folding time $\tau_\text{FP}(q_1,
  q_1^f)$ for the final state $q_1^f=0.11$, as extracted directly from
  the simulation data (circles) and compared to predictions from
  Eq.~(\ref{eq:2}) using the different diffusivities shown in (d).
  (e) MFP or unfolding time for the final state $q_1^f = 0.57$, same
  notation as in (d).  Vertical dotted lines in (d) and (e) mark the
  final states $q_1^f$ for folding and unfolding.}
\label{fig3}
\end{figure}

In Fig.~3 we focus on RC $q_1$.  The free energy profile $\beta F(q_1)
= - \ln \langle \Psi (q_1) \rangle$ in a) reveals the intermediate
state and two barriers at $q_1 \approx 0.26$ and $q_1 \approx 0.48$.
Fig.~3(b) shows the roundtrip times $\tau_\text{RT}(q_1, q_1^f)$ for
various final states $q_1^f$ as a function of $q_1$, directly
extracted from the simulation time series~\cite{Supplement}.  The data
sets are shifted vertically (which according to Eq.~(\ref{eq:5}) is
irrelevant for extracting $D(q_1)$) to illustrate the predicted
collapse onto a single mean round-trip time function
$\bar{\tau}_\text{RT}(q_1)$.  The smooth fit
$\bar{\tau}_\text{RT,fit}(q_1)$ is shown as a blue curve.  The
collapse of $\tau_\text{RT}(q_1, q_1^f)$ for different $q_1^f$ is a
strong check on the consistency of the FP approach.  The red curve
denotes the round-trip time from the Bayesian approach~\cite{Hummer},
obtained for optimized time interval and smoothing parameters $\Delta
t=6~\text{ns}$ and $\gamma=0.2~\text{ns}^{-1}$~\cite{Supplement}.
Fig.~3(c) shows the diffusivity $D(q_1)$ extracted from
$\bar{\tau}_\text{RT,fit}(q_1)$ via Eq.~(\ref{eq:5}) (blue curve).
Most notably, $D(q_1)$ varies considerably along $q_1$: it is reduced
by an order of magnitude around the intermediate state at $q_1 \approx
0.32$ and seems correlated with $F(q_1)$.  The $D(q_1)$ profile from
the Bayesian approach (red curve) reproduces the coarse features of
our round-trip approach with slight difference that will be discussed
below.  We stress that we have fitted the two parameters in the
Bayesian approach, namely the time interval and the smoothing
parameter, by a comparison with the simulation mean-first passage
times (see Supplement for further details~\cite{Supplement}). The
diffusivity profiles resulting from the Bayesian approach sensitively
depend on these parameters, and without such a comparison it is not
easy to see what are sensible parameter values. This highlights an
advantage of our method based on the round-trip time, since the only
parameter is a smoothing factor that operates directly on the
round-trip time, a physical observable, and sensible parameter values
are straightforwardly estimated.  The variance method Eq.~(\ref{eq:6})
for lag time $\delta t=200$~fs (upper green curve) overestimates
$D(q_1)$ by two orders of magnitude, yet for $\delta t=200$~ps (lower
green curve) $D_\text{var}$ approaches the results of the other two
methods quite nicely.  Thus for $\delta t < 200$~ps, $D_\text{var}$ is
dominated by non-Markovian events that are unrelated to the long-time
folding/unfolding dynamics; interestingly, this threshold time is
similar to the transition time for helix unwrapping inferred from
Fig.~1(d).  In Figs.~3(d) and (e), we show MFP times
$\tau_\text{FP}(q_1,q_1^f)$ for $q_1>q_1^f=0.11$ (folding) and
$q_1<q_1^f=0.57$ (unfolding) calculated from Eq.~(\ref{eq:2}) and the
various $D(q_1)$ profiles shown in (c).  $\tau_\text{FP}(q_1,q_1^f)$
directly extracted from simulation data (circles) in Fig.~3(d) is most
accurately reproduced by the Bayesian fitting approach (red curve), as
expected since the probability distribution and thus the frequency of
transitions is maximal in the range $q_1 \approx 0 - 0.25$ (see
Fig.~2(a). The RT approach (blue curve) considers an equal balance of
folding and unfolding events and consequently describes unfolding MFP
times in Fig.~3(e) better.  Noteworthy, the RT approach is simple to
implement, directly works on the property one wishes to describe
(namely folding/unfolding times) and has apart from the functional
form of the fitted round-trip time $\bar{\tau}_\text{RT}(q_1)$ no
freely adjustable parameter.  The combined deviations between
simulation data and Fokker-Planck predictions in Figs.~3(d,e) are due to
a combination of non-Markovian processes at short times and
insufficient trajectory sampling.

\begin{figure}[t!]
\centering
\includegraphics*[width = 7.2cm]{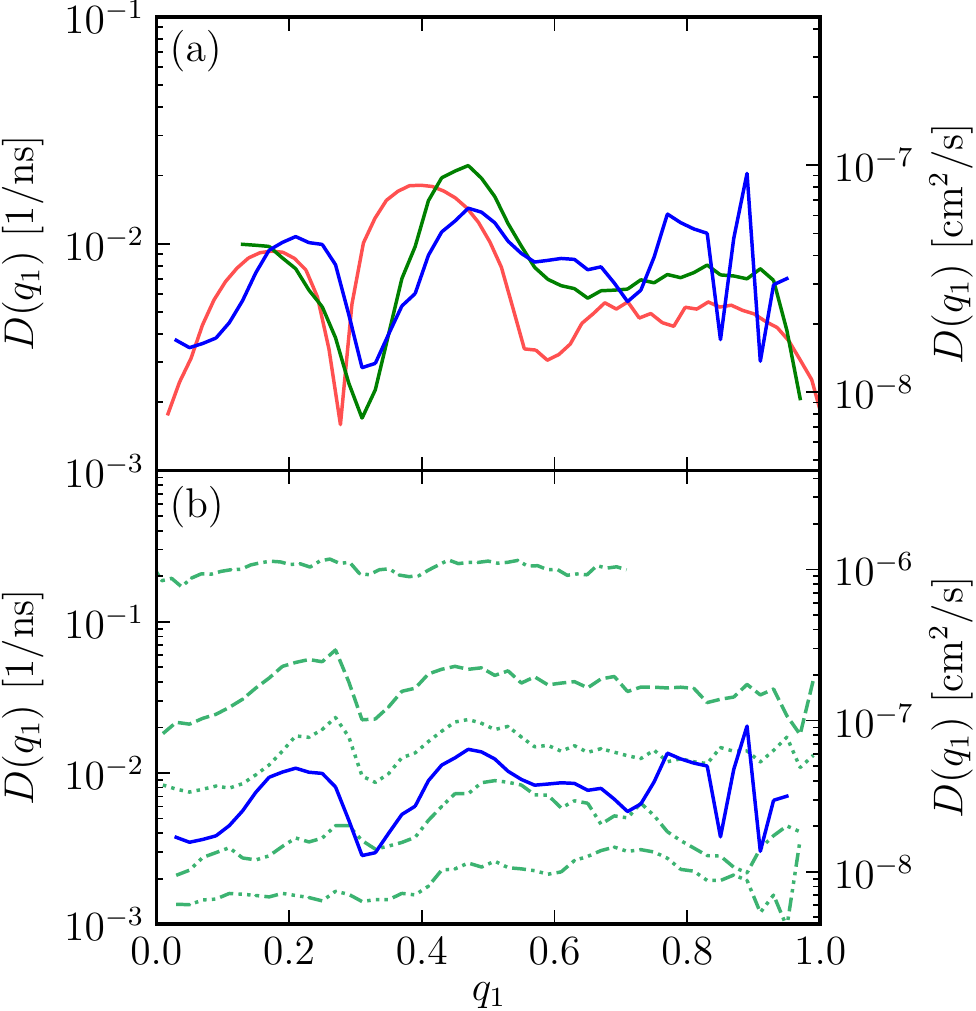}
\caption{ Results for RC $q_1$.  (a) Diffusivity from round-trip time
  method Eq.~(\ref{eq:5}) (blue curve) and the Bayesian approach (red
  curve); these are the same data already shown in Fig.~3(c). The
  green curve is based on the first passage time method and follows
  from Eq.~(\ref{eq:3}) for the final state $q_1^f =0.11$.  (b)
  Diffusivity from the round-trip time method Eq.~(\ref{eq:5}) (blue
  curve) compared to the variance method Eq.~(\ref{eq:6}) for lag
  times $\delta t= 200$~fs, 20~ps, 200~ps, 2~ns and 10~ns (green
  curves, from top to bottom).}
\label{fig4}
\end{figure}

In Fig.~4(a) we compare the diffusivities based on the round-trip time
approach (blue curve) and the Bayesian approach (red curve), already
presented in Fig.~3(c), with results obtained from the MFP times via
Eq.~(\ref{eq:3}), shown as a green curve.  For the fit we used a final
state $q_1^f =0.11$ and considered folding events from $q_1>q_1^f $ to
$q_1^f $. It is seen that the three curves roughly coincide, which
testifies to the robustness of methods for deriving diffusivities from
folding times.  In Fig.~4(b) we compare diffusivities from the
variance method, Eq.~(\ref{eq:6}), to the round-trip time method
Eq.~(\ref{eq:5}) (blue curve). Here we present results for
$D_\text{var}(Q, \delta t)$ for a wider range of lag times of $\delta
t= 200$~fs, 20~ps, 200~ps, 2~ns and 10~ns (green curves, from top to
bottom).  It is seen that for lag times between $\delta t= 200$~ps and
$\delta t= 2$~ns, $D_\text{var}(Q, \delta t)$ agrees with the
round-trip time approach.  As already discussed, for smaller lag times
$D_\text{var}(Q, \delta t)$ is too large.  For larger lag times
$D_\text{var}(Q, \delta t)$ loses structure and becomes too small,
which has to do with the fact that at those times the peptides
explores a considerable subsection of the free energy space and the
effect of the energetic barriers encountered are spuriously accounted
for by a reduction of the diffusivity. The situation is similar to the
Bayesian approach: there is no a-priori way of knowing what the
suitable parameter value for the lag time is, unless one compares to a
physical observable, which might be the folding or round-trip time. In
that case, however, a direct fitting of $D(q)$ based on folding times
as suggested by us seems more direct and transparent.

\begin{figure}[t!]
\centering
\includegraphics*[width = 7.2cm]{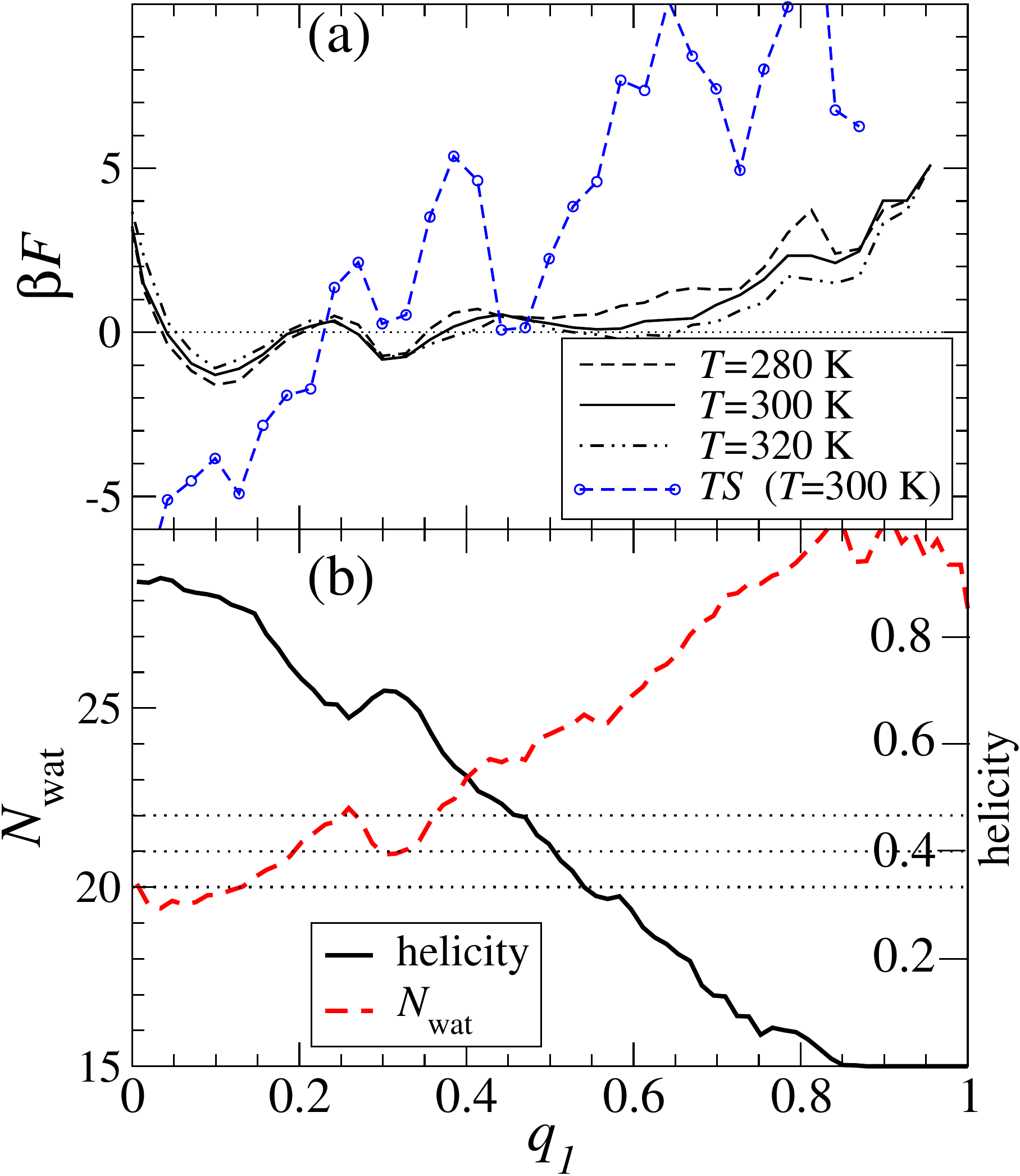}
\caption{ (a) Replica-exchange MD results for the free energy profile
  $\beta F(q_1)$ for different temperatures $T$, together with the
  entropic contribution $TS$ obtained from the finite-$T$ difference
  (with $\Delta T=20$~K) of $\beta F(q_1)$.  (b) Helicity and the
  number $N_\text{wat}$ of backbone-bound water molecules vs. $q_1$ at
  $T=300K$.  }
\label{fig5}
\end{figure}
\begin{figure}[t!]
\centering
\includegraphics*[width = 7.2cm]{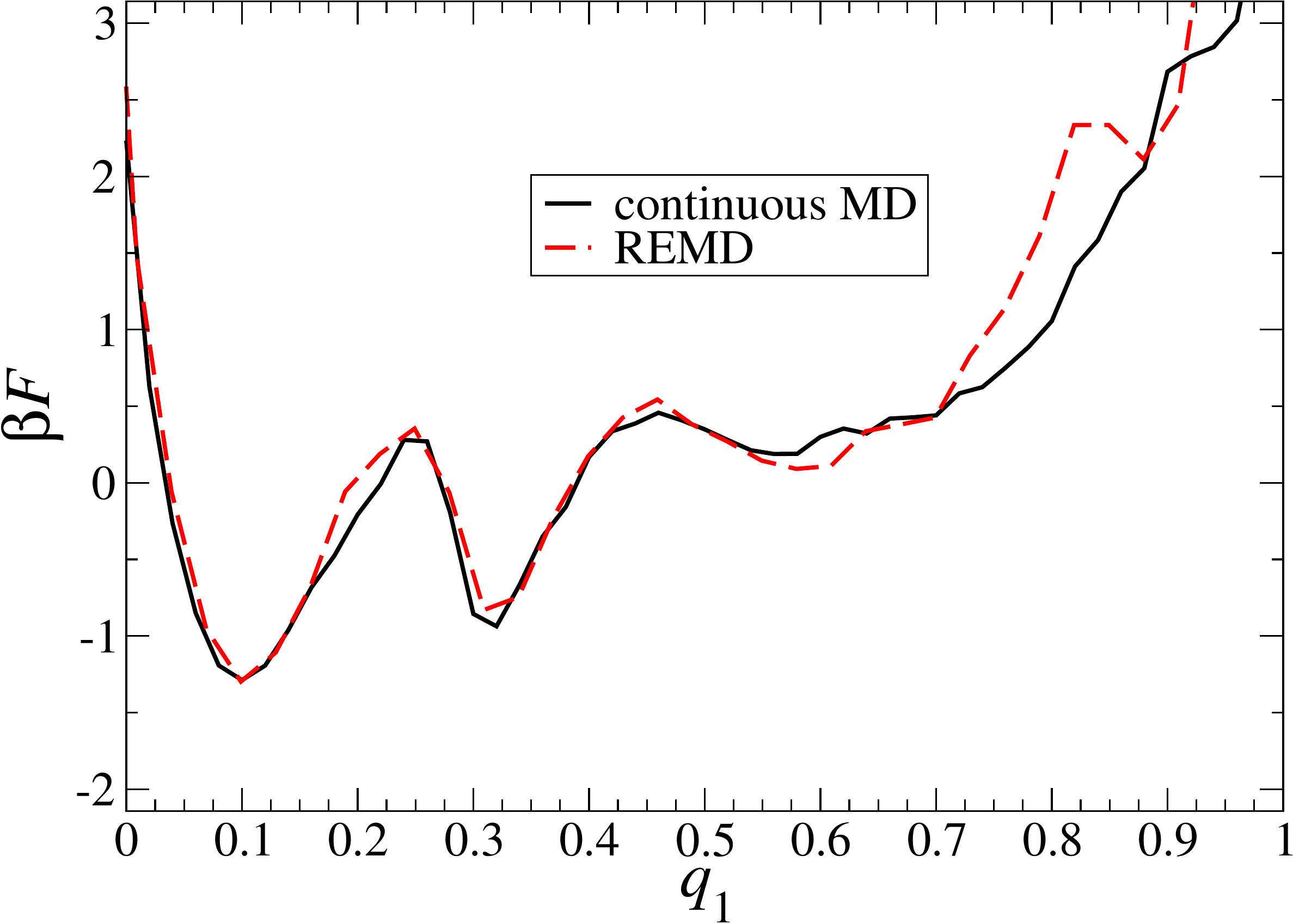}
\caption{ Comparison between replica-exchange MD results (red broken curve)
and standard MD results (black solid curve)  for the free energy profile $\beta F(q_1)$  at $T=300K$.
}
\label{fig6}
\end{figure}
\begin{figure*}[t!]
\centering
\includegraphics*[width = \textwidth]{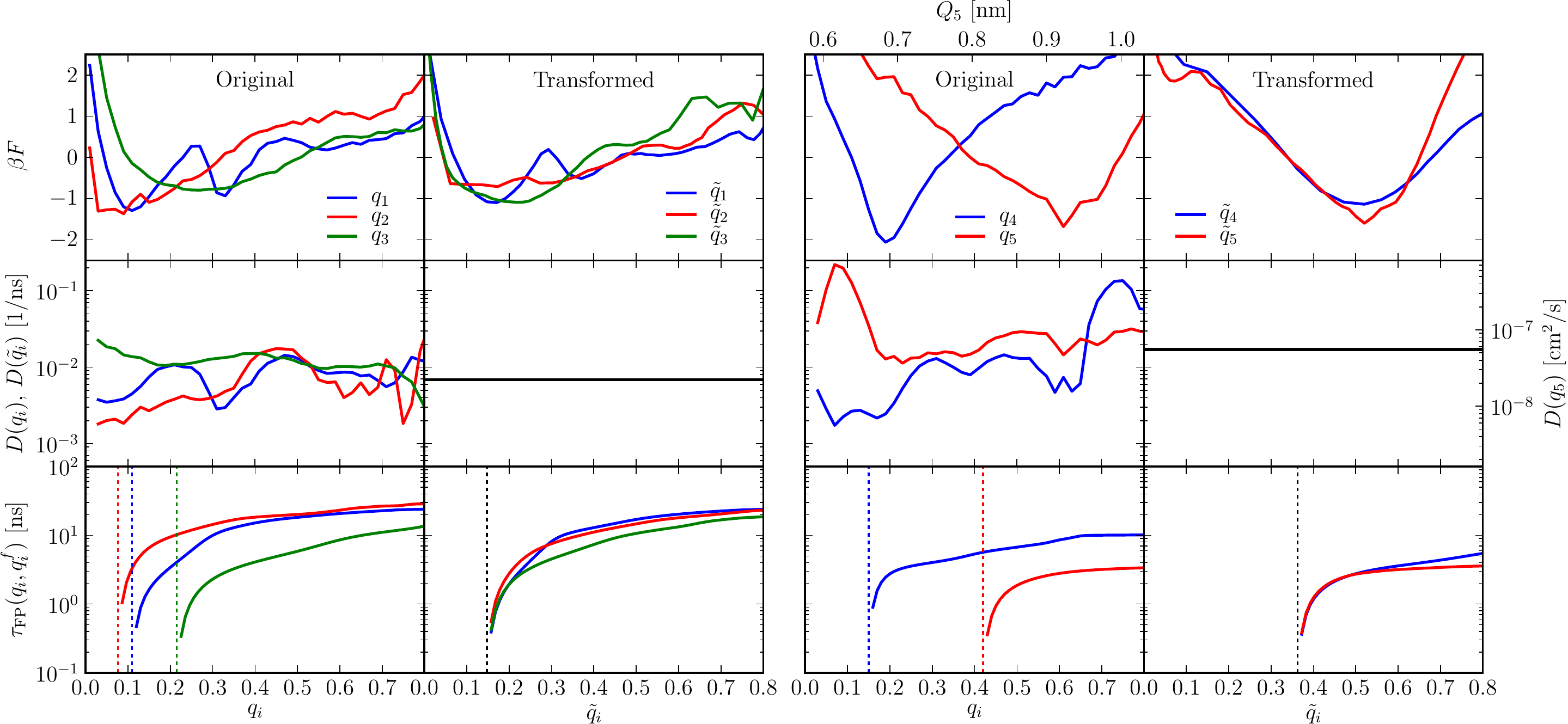}
\caption{ Free energy profiles (top row), diffusivity profiles (middle
  row), and folding MFP times (bottom row) for all five reaction
  coordinates. The columns denoted ``Original'' show results as a
  function of the original RCs $q_i$, while in ``Transformed''
  rescaled coordinates $\tilde{q_i}$ are used such that the
  diffusivity profiles are constant.  The final states $q_i^f$ for the
  folding (marked by dotted vertical lines) are chosen such that they
  map onto a single value $\tilde{q}_i^f$ separately for the
  $q_1,q_2,q_3$ and $q_4,q_5$ groups.}
\label{fig7}
\end{figure*}

A free energy barrier, as exhibited by $F(q_1)$ in Fig.~3(a), was
argued to arise from a subtle compensation of energy and entropy
effects, which both increase upon unfolding~\cite{Wolynes}.  This
scenario, developed in the context of lattice models, is basically
confirmed by our explicit water simulations.  In Fig.~5(a), we show free
energy profiles at different temperatures $T$ from replica-exchange
simulations.  Indeed, the entropic contribution $TS$, estimated from
the free energy difference between $T=$280K and 320K, shows
considerable numerical error but rises across the unfolding
transition.  In Fig.~5(b) we show the number $N_\text{wat}$ of
backbone-bound water molecules that have a distance to a backbone
oxygen smaller than 0.35nm.  Apart from the loss of one bound water
molecule at $q_1 \approx 0.3$ (paralleled by a helicity increase),
$N_\text{wat}$ steadily rises from about $N=20$ in the folded state to
$N=30$ in the unfolded state.  So we conclude that the entropy
increase upon unfolding results from a competition of water binding
and conformational effects.  The overall good comparison between the
free energy profile from a standard MD simulation run (for a length of
1.1$\mu$s) and results from a replica exchange MD simulation
(trajectory length 22.5 ns and equilibrated with 32 replicas at
different temperatures) at T=300K in Fig.~6 gives good evidence that
the times series considered in our kinetic analysis is long enough.

The appearance of a free energy barrier, as seen in $F(q_1)$ in
Fig.~3(a), is often interpreted as equivalent to exponential kinetics,
which is not necessarily true as we will now discuss.  In fact, even
the presence of a free energy barrier depends on the specific RC
employed and thus is a much less robust feature than often assumed: In
Fig.~7 we show the free energy $F(q_i)$ and diffusivity $D(q_i)$
profiles of all five RCs. We separate RCs that embody knowledge of the
native state $q_1, q_2, q_3$ and the unbiased RCs $q_4, q_5$. In the
columns ''Original'' we use the bare RCs $q_i$ as defined in the
Methods section, in the columns ''Transformed'' we use rescaled RCs
$\tilde{q}_i$ such that the diffusivities are constant,
$\tilde{D}(q_i) = \tilde{D}_0$.  Two features strike the eye: 

i) Most
diffusivity profiles are full of structure and vary substantially
along the reaction path; it immediately transpires that a description
of the folding kinetics without consideration of the diffusivity
profile can fail.  

ii) The profiles $F(q_i)$ and $D(q_i)$ vary
considerably among different RCs. In fact, while $F(q_1)$ shows
pronounced barriers and an intermediate state, the profiles $F(q_2)$
and $F(q_3)$ are free of barriers: We conclude that the presence of
barriers depends on the RC chosen.  Do the kinetics within an
effective Fokker-Planck description also vary among RCs, possibly
showing exponential for some and non-exponential behavior for other
RCs?  While the free energy profiles $F(q_i)$ as a function of the
original RCs show large variations, the profiles
$\tilde{F}(\tilde{q}_i)$ after the transformation are quite similar
(this is most striking for the radius of gyration, $\tilde{q}_4$, and
the end-to-end radius, $\tilde{q}_5$), and thus the kinetics as
characterized by the MFP times $\tau_\text{FP}(q_i,q_i^f)$ in the
bottom row are very similar.  This at first surprising result can be
easily rationalized: the round-trip method is designed to optimally
reproduce the complete set of round-trip times and thus the slowest
conformational transitions in the system.  The different diffusivities
$D(q_i)$ and free energy profiles $F(q_i)$ together uniquely determine
the folding times. Assuming that different RCs yield a comparable
separation of states into the unfolded and folded basins, it follows
that the folding times must be very similar. This in fact holds for
the RCs $q_1, q_2, q_3$ on the one hand and for the RCs $q_4, q_5$ on
the other hand.  Since after the rescaling the entire kinetic
information is contained in the free energy profile, those profiles
must be quite similar.  It follows that the presence of a free energy
barrier does not necessarily imply exponential kinetics; for that
statement to be true the free energy barrier must persist after a RC
transformation that makes the diffusivity profile flat.  Although
there are still differences among the free energy profiles for $q_1,
q_2, q_3$ after the transformation, they are small enough that the
kinetics are not particularly distinguished.

\begin{figure}[t!]
\centering
\includegraphics*[width=7.2cm]{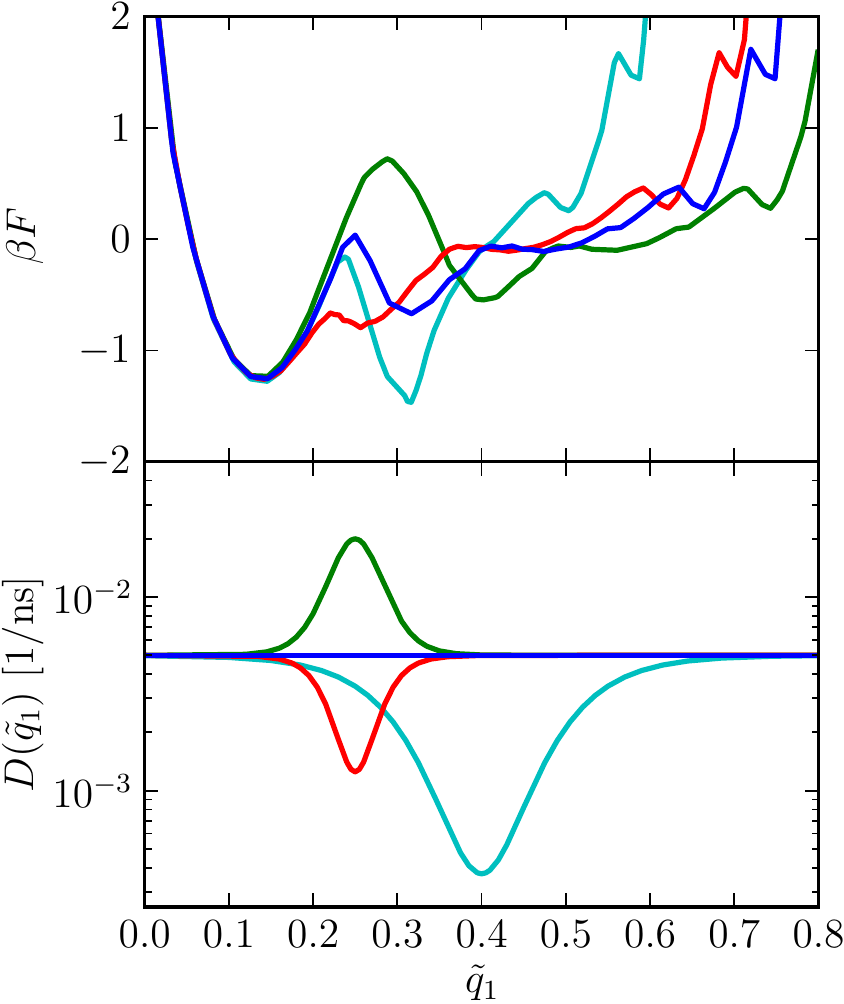}
\caption{ Free energy (top) and diffusivity profiles (bottom) for
  different rescaled RCs $\tilde{q}_1$ .  Starting from the RC
  exhibiting a flat diffusivity (shown in blue), we arbitrarily rescale $\tilde{q}_1$
  according to Eq.~( \ref{fantasy}) such as to increase the barrier
  (green), decrease the barrier (red) and to reolocate the stable
  minimum (turquoise).  }
\label{fig8}
\end{figure}

To highlight the implications of these results,
we now turn the argumentation around. 
Consider a general RC  transformation 
\begin{equation} \label{fantasy}
\tilde{q}  = q + c (  \text{Tanh} [(q-q^*)/ d]-1),
\end{equation}
that is assumed to be a monotonic function which implies that $d>-c$. This rescaling
corresponds to a local stretching / compression of the RC around $q^*$  and 
via the reparametrization properties of the Fokker-Planck equation
also modifies the diffusivity and the free energy profiles. 
In Fig.~8 we show three different rescaled $\tilde{F}(\tilde{q}_1)$ and $\tilde{D}(\tilde{q}_1)$ profiles, 
all generated via Eq.~(\ref{fantasy}) from the  RC $\tilde{q}_1$ for  which $\tilde{D}(\tilde{q}_1)$ is flat (shown in blue). 
Depending on the parameters $q^*, c, d$ we generate free energy profiles that either exhibit a more pronounced
barrier (green curve), a reduced barrier (red curve), or a free energy profile where the position of the minimum is moved
from the folded to the unfolded state (turquoise curve). We mention that by construction, the kinetics as characterized
by  the round trip or MFP time
are invariant under this rescaling. What this figure demonstrates is that under a combined rescaling of $F(q)$ and
$D(q)$ one can generate a bewildering variety of free energy curves which share the identical kinetics,
meaning that the free energy profile without the diffusivity is not sufficient to even qualitatively
predict protein folding kinetics.

\begin{figure}[t!]
\centering
\includegraphics*[width = 8.2cm]{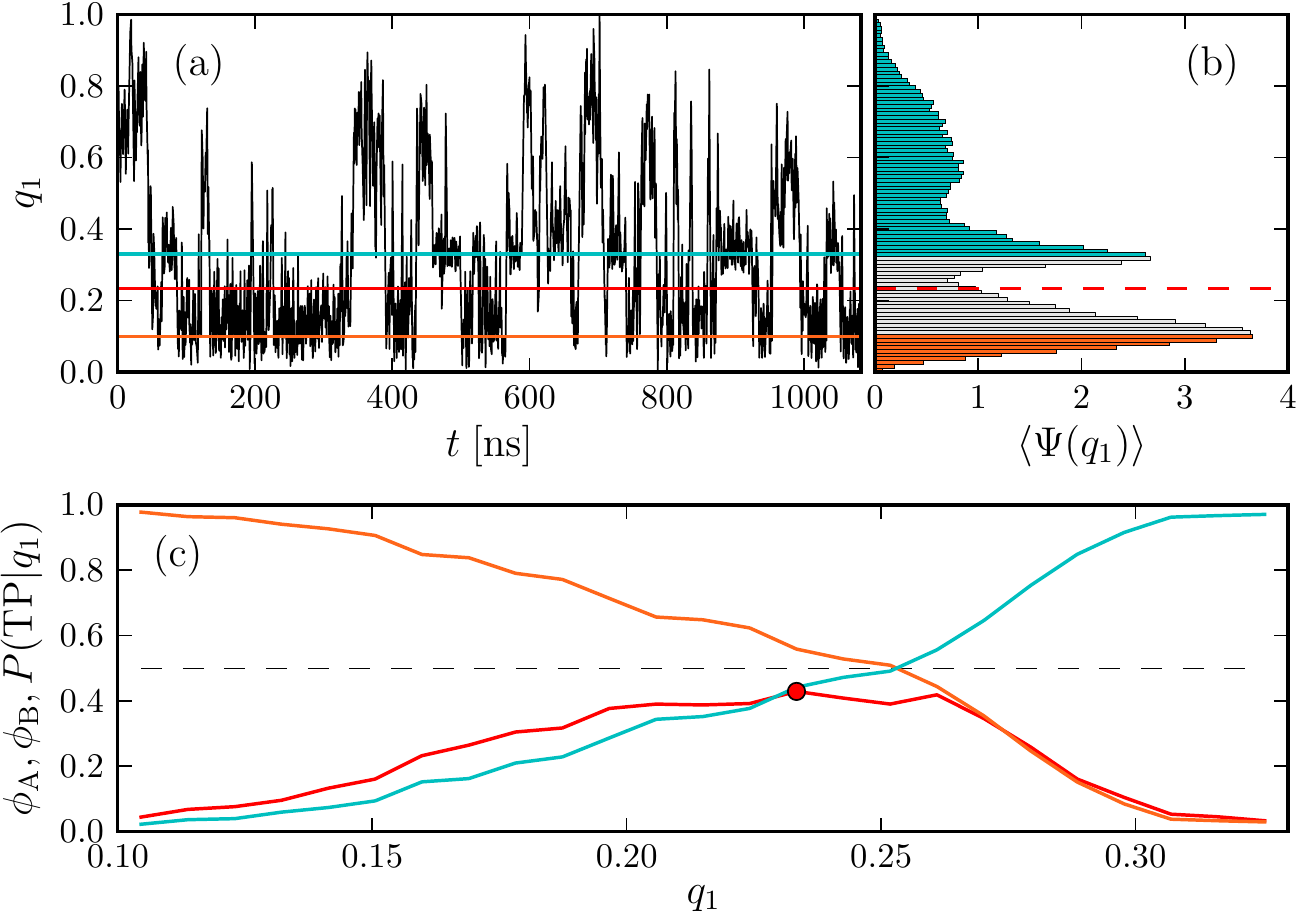}
\caption{ Test for the quality of reaction coordinate $q_1$. (a) shows
  the complete trajectory, (b) shows the corresponding equilibrium
  distribution $\langle \psi(q_1) \rangle $ and the regions A
  ($q_1<0.1$) and B ($q_1>0.33$) marked in orange and blue,
  respectively.  The complete trajectory contains 181 transitions
  between A and B (90 from A to B and 91 from B to A). Panel (c) shows
  the splitting probabilities $\phi^\text{A}(q_1)$ (orange) and
  $\phi^\text{B}(q_1)$ (blue) and the transition path probability
  $P(\text{TP} | q_1)$ (red). $P(\text{TP} | q_1)$ reaches the maximum
  value $P(\text{TP} | q_1) \approx 0.43$ for
  $q_1=q_1^\ddagger\approx0.23$, denoted by a red circle in (c) and
  red lines in (a) and (b).  }
\label{fig9}
\end{figure}

Much of the discussion in the preceding sections and the usage of
one-dimensional RCs presumes that the reaction coordinates are
``good'' in the sense that i) the ensemble of transition states is
assigned to a narrow region of RC values and ii) that the probability
of finding a transition state in that region is
maximal~\cite{Bolhuis,BestHummer2}.  To make that notion more
concrete, one introduces the splitting probabilities
$\phi^\text{A}(q)$ and $\phi^\text{B}(q)$ for each value of the RC,
where $\phi^\text{A}(q)$ is the probability to reach, starting from RC
value $q$, region A before region B~\cite{BestHummer2}.  In the
context of transition states, the regions A and B would denote regions
corresponding to the folded and unfolded domains flanking the
transition region. The splitting probabilities are normalized as
\begin{equation}
\phi^\text{A}(q) + \phi^\text{B}(q)=1
\end{equation}
since eventually any state will diffuse out towards the boundaries. 
For a trajectory that passes through state $q_1$ there are four choices, it can be
trajectory starting in A and returning to A, starting in B and returning B,
starting in A and ending up in B or starting in B and ending in A. The 
respective probabilities are normalized as
\begin{equation}
\begin{split}
&P(\text{A} \rightarrow \text{A} | q) +
P(\text{A} \rightarrow \text{B} | q) \\
&\quad + P(\text{B} \rightarrow \text{A} | q) +
P(\text{B} \rightarrow \text{B} | q) =1.
\end{split}
\end{equation}
For non-ballistic stochastic motion, the transition path probability
$P(\text{TP} | q) = P(\text{A} \rightarrow \text{B} | q) + P(\text{B}
\rightarrow \text{A} | q)$, i.e. the probability that the trajectory
connect regions A and B, can be maximally 1/2. A maximum close to 1/2
characterizes a good reaction coordinate, a significantly smaller
number points to a bad reaction coordinate. In Fig.~9 we show a
detailed reaction coordinate analysis for RC $q_1$ with a resolution
of 25 bins in the range $0.1 < q_1 < 0.33$ and using the full time
resolution of 20~ps. In (a) we show again the complete time series and
in (b) the corresponding probability distribution. Region A for
$q_1<0.1$ is the folded region, region B for $q_1> 0.33$ is a region
where one helical turn is unfolded.  In (c) we show the splitting
probabilities $\phi^\text{A}(q_1)$ and $\phi^\text{B}(q_1)$ (orange
and blue lines). The behavior is as expected, with the probabilities
switching from zero to unity between the boundaries of the regions A
and B, and a rather large slope in the region around $q_1 \approx 0.25
- 0.30$.  The maximum of the transition path probability $P(\text{TP}
| q_1^\ddagger) \approx 0.43$ (shown as a red curve) at a position
$q_1^\ddagger\approx0.23$ means that $q_1$ is quite close to a perfect
reaction coordinate and that the Fokker-Planck analysis performed in
this paper is appropriate for long times on the order of folding and
unfolding events. Note that  $q_1^\ddagger\approx0.23$  is close to a minimum in the
equilibrium  distribution $\langle \psi(q_1) \rangle $, see Fig.9b, at which position
the free energy thus exhibits a maximum.
This is coincidental, since as we have shown in Fig.8, one can easily change
the free energy profile by a reaction-coordinate rescaling, which however leaves the
splitting probabilities and the transition path probabilities invariant.

\section{Conclusions}

In the naive approach towards protein kinetics, folding times are
deduced from the free energy profile $F(Q)$ alone.  As has been argued
before,~\cite{BestHummer,Wang,BestHummer3,Onuchic} such an approach is
unreliable since for the simplest non-trivial folder, namely a single
short $\alpha$-helix in explicit solvent, the diffusivity profile
$D(Q)$ varies substantially along the folding path.  Our $D(Q)$
variation comes out somewhat stronger than from similar simulations
with implicit solvent, suggesting that explicit solvent further
increases the importance of diffusivity
inhomogeneities~\cite{BestHummer}.
In fact, to match experimental folding times of simple alpha-helix forming
oligo-peptides within solvent-implicit simulations, an overall correction factor 
to the time scales is typically applied\cite{Chowdhury,Wang2}. 
A detailed microsopic justification for this is lacking; on the contrary, it has
been shown that in many cases explicit solvent strongly influences
the free energy landscape and introduces 
novel kinetic mechanisms that are completely absent in solvent-implicit 
simulations\cite{Zhou,Pande2}.
When extending the analysis to five
different popular reaction coordinates, we find free energy and
diffusivity profiles to vary substantially among different RC
representations.  Yet, the kinetics that follows from a Fokker-Planck
description is largely independent of the RC chosen, if and only if
$D(Q)$ is properly accounted for.  A similar conclusion was reached
recently based on coarse-grained, solvent-implicit
simulations~\cite{BestHummer3}.  This means that a quasi-universal
(i.e. RC independent) description of protein folding kinetics
necessarily involves $D(Q)$.  For this quasi-universality to hold we
have to distinguish between reaction coordinates that are based on the
distance to the native state (such as $Q_1, Q_2, Q_3$) and those that
are purely geometric in nature (such as $Q_4, Q_5$).  By considering
generalized RCs and using the reparametrization invariance of the
Fokker-Planck equation, we can design arbitrary $F(Q)$ profiles with
no barrier at all, an enhanced barrier, or an interchange of the naive
stable and unstable states. This means that the concept of a free
energy profile is to some degree arbitrary, which might be relevant
with regards to recent discussions in the experimental
literature~\cite{Gruebele,Munoz,Fersht2}.  The kinetics, embodied in
the folding time, and dependent on $F(Q)$ and $D(Q)$, is less
arbitrary.

Our simulations are for a single $\alpha$-helix fragment, one of the
shortest oligopeptides which shows non-trivial folding.  There is no
reason to believe that for larger proteins the situation will
simplify; we therefore argue that the diffusivity profile will be full
of features and thus important in those more complicated situations as
well.  Our conclusions also apply to optimized or otherwise carefully
selected RCs~\cite{Bolhuis,BestHummer2,Eric,Orland,Noe}, since the
reparametrization can be done for any RC and thus arbitrarily create,
annihilate and shift barriers in the folding landscape (incidentally,
RC $q_1$ turns out to be a quite good reaction coordinate according to
the definition of Ref.~\cite{BestHummer2}, as shown in Fig.~9).  Our
method of extracting the diffusivity profile via the mean-first-passage or
round-trip  time formalism can be easily applied to time series
data from FRET or force-spectroscopic experiments, 
so an experimental test of our results is possible.


\section{Acknowledgements}
We acknowledge support from the Deutsche Forschungsgemeinschaft (DFG)
within SFB 863 and the Emmy-Noether-Programme (JD). 
The Leibniz Rechenzentrum (LRZ) Munich is acknowledged for supercomputing access.


\newpage

\renewcommand{\theequation}{S.\arabic{equation}}
\renewcommand{\thefigure}{S.\arabic{figure}}
\renewcommand{\thesection}{S.\arabic{section}}
\setcounter{equation}{0}
\setcounter{figure}{0}
\setcounter{section}{0}

{\centering \bf \large Supplement to: ``How the diffusivity profile reduces the arbitrariness of protein folding free energies''\\}

\begin{center} M. Hinczewski, Y. von Hansen, J. Dzubiella, R. R. Netz
\end{center}

\section{Mapping between reaction coordinates}

Fig. 1 shows in the columns the average distribution function $\langle
\Psi (q_i) \rangle$ as a function of all five different RCs considered
in the main text.  In each row the coloured regions denote identical
subsets of states and are chosen to correspond to pure states for one
reaction coordinate.  While among the RCs $q_1, q_2, q_3$ and among
the RCs $q_4, q_5$ the ordering of the coloured regions is preserved,
this ordering is lost between those two groups.  This points to a
fundamental difference between the RCs $q_1, q_2, q_3$, that embody
knowledge of the native state, and the RCs $q_4, q_5$, which are
purely geometric.
\begin{figure*}[t]
\centering
\includegraphics[width = 15cm]{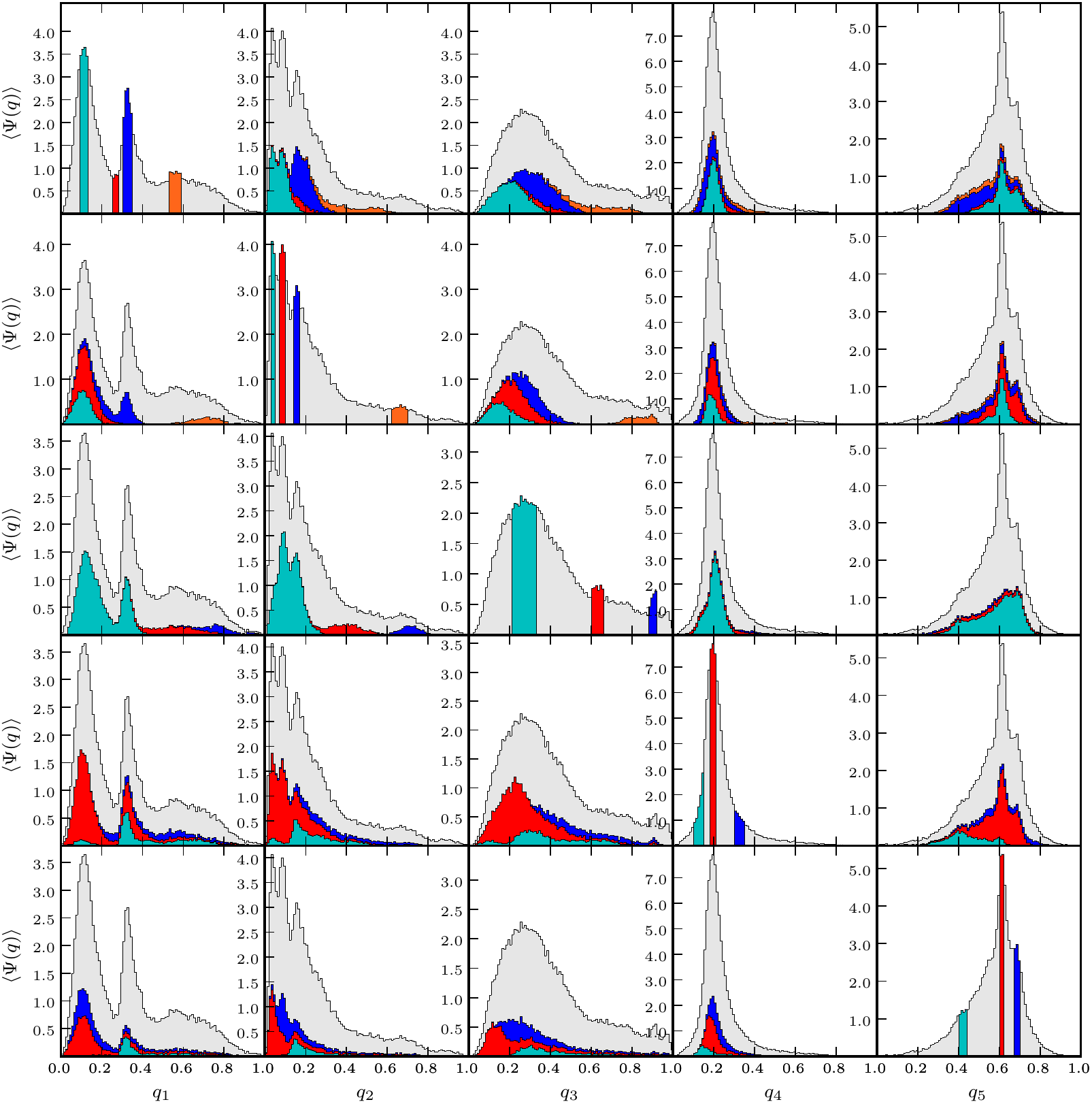}
\caption{
Mapping between different reaction coordinates.
Columns show the density distribution 
$\langle \Psi (q) \rangle$ plotted in terms of the different reaction coordinates
$q_1$, $q_2$, $q_3$, $q_4$, and $q_5$.
In each row selected regions of the distribution are shown in color.
 }
\label{sfig1}
\end{figure*}

\section{Extracting the diffusivity profile}

In Fig. 2 we show the free energy profiles, the round-trip times and
the diffusivity profiles of all five reaction coordinates.  In Fig. 3
we show the mean-first passage times for folding and unfolding events
for all reaction coordinates, as extracted from the fitted diffusivity
profiles and the Fokker-Planck description.  The final states $q_i^f$
were chosen such that $\approx 20\%$ of the probability distribution
$\langle \Psi(q_i) \rangle$ is contained in the range $0\le q_i \le
q_i^f$ (folding), or $q_i^f \le q_i \le 1$ (unfolding).  The noise and
non-monotonicity in the $\tau_\text{FP}$ curves extracted from the
simulation data are due to the statistical effects of insufficient
trajectory sampling (particularly at the edges of the free energy
landscape) and time discretization.
\begin{figure*}[t]
\centering
\includegraphics[width = 15cm]{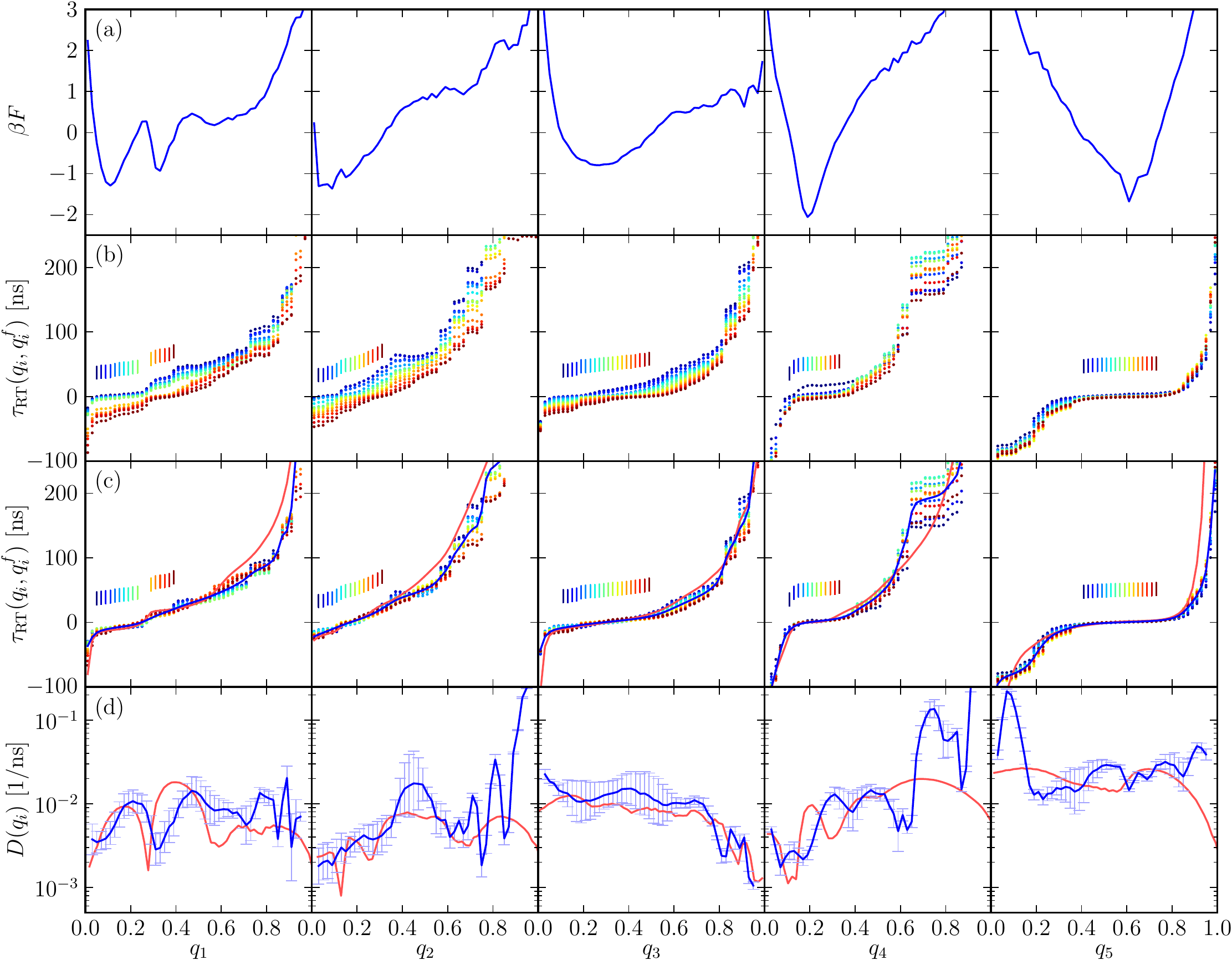}
\caption{The columns give results for all different RCs considered.
  a) Free energy profile $\beta F (q_i) = - \ln \langle \Psi(q_i)
  \rangle$.  b) Data points give the round-trip times
  $\tau_\text{RT}(q_i, q_i^f)$ as extracted from the simulation data
  for various final states $q_i^f$ that are denoted by vertical
  colored bars.  c) Same data shifted vertically to illustrate the
  approximate collapse onto a single mean round-trip function
  $\bar{\tau}_\text{RT}(q_i)$ for all $q_i^f$, with the smooth fit
  function $\bar{\tau}_\text{RT,fit}(q_i)$ shown as a blue curve.  The
  red curve denotes the round-trip time from the Bayesian approach.
  d) Diffusivity from the round-trip time method (blue curve), compared to
  the Bayesian method (red curve).  }
\label{sfig2}
\end{figure*}
\begin{figure*}[t]
\centering
\includegraphics[width = 15cm]{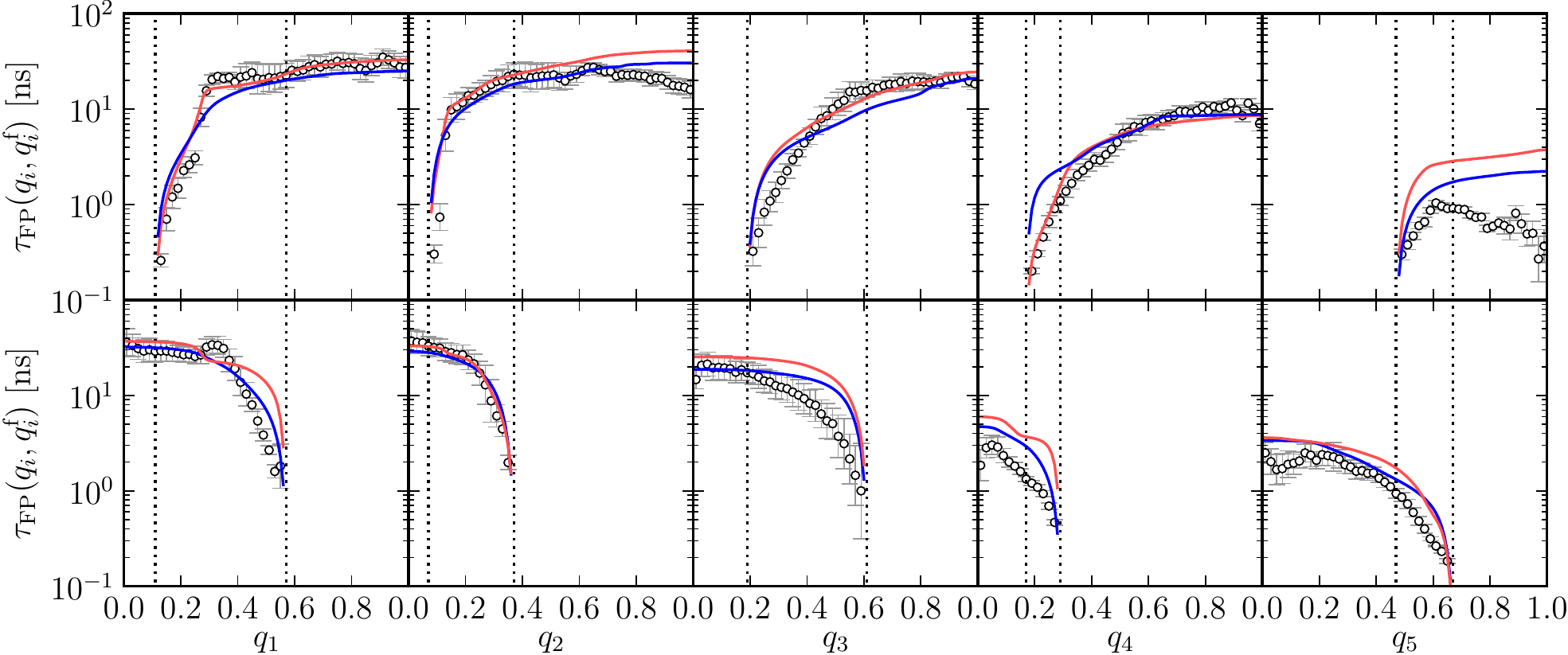}
\caption{The columns give results for all different RCs
  considered. Upper row: MFP or folding time $\tau_\text{FP}(q_i,
  q_i^f)$ for different final states $q_i^f$, as extracted directly
  from the simulation data (circles) and compared to Fokker-Planck
  predictions using the diffusivities from the round-trip time
  approach (blue curves) and the Bayesian approach (red curves). The
  optimized parameters of the Bayesian approach are $\gamma =
  0.2/\text{ns}$ and $\Delta t =6$~ns for $q_1$ and $q_2$ and
  $\Delta t =2$~ns for $q_3$, $q_4$ and $q_5$.  Lower row: MFP or
  unfolding time, same notation as in upper row.  The vertical dotted
  lines in both rows mark the final states $q_i^f$ for folding and
  unfolding.}
\label{sfig3}
\end{figure*}

\section{Determining Diffusivity Profiles by Bayesian Inference}
We briefly review the optimization method introduced in
Ref.~\cite{Hummer2005}, and used previously to extract diffusivity
profiles for protein folding dynamics in implicit
solvent~\cite{Best2006}.

\subsection{Master equation approach}
When discretized in reaction coordinate space, the FP equation takes the form of a master equation~\cite{Bicout1998}
\begin{equation}
\label{Eq:MasterEqFP}
 \frac{\partial \Psi_i(t)}{\partial t}=R_{i,i-1}\Psi_{i-1}(t)+R_{i,i+1}\Psi_{i+1}(t)-R_{i,i}\Psi_i(t),
\end{equation}
where the probability of being in bin $i$ is denoted by $\Psi_i(t)\equiv\Psi(Q^{(i)},t)\Delta Q$, the bin width is $\Delta Q$, the bin index $i$ ranges from $1$ to $M$, and the transition rate from bin $j$ to bin $i$ is $R_{i,j}$. The rates fulfill detailed balance, i.e. $R_{i,j}\Av{\Psi_j}=R_{j,i}\Av{\Psi_i}$, where the equilibrium probability of each bin $i$ is denoted by $\Av{\Psi_i}$; the loss in bin $i$ is caused by transitions to neighboring bins, i.e. $R_{i,i}=-\sum_{j\neq i}R_{j,i}$. The rates in the master equation~\ref{Eq:MasterEqFP} are related to the free energy $F(Q)$ and the diffusivity profile $D(Q)$ in the FP equation via:
\begin{equation}
\label{Eq:RelationMasterFP}
 F(Q^{(i)})\approx-k_\text{B}T\log{\left(\frac{\Av{\Psi_i}}{\Delta Q}\right)},
 \end{equation}
\begin{equation}
D_{i+1/2}\approx(\Delta Q)^2R_{i,i+1}\sqrt{\frac{\Av{\Psi_{i+1}}}{\Av{\Psi_i}}},
\end{equation}
with $D_{i+1/2}\equiv(D(Q^{(i)})+D(Q^{(i+1)}))/2$ being the diffusivity between the bins. For $M$ bins the system is consequently characterized by $2M-1$ independent parameters: $M-1$ rates $R_{i,i+1}$ for transitions from the neighboring bin on the right hand side and $M$ equilibrium probabilities $\Av{\Psi_i}$.

\subsection{Bayesian analysis of trajectories}
In a system described by Eq.~\ref{Eq:MasterEqFP} the conditional probability of landing in bin $i$ in time $\Delta t$ given a start in bin $j$ is: 
\begin{equation}
\label{Eq:GFMasterEq}
p(i\vert j;\Delta t)=\left(\exp{\left(\Delta t\T R \right)}\right)_{i,j},
\end{equation}
where $\T R$ is the matrix with entries $R_{i,j}$. In our case $\T R$
is tridiagonal and the transition probabilities are easily obtained
numerically by diagonalization of the symmetrized matrix $ \tilde {\T
  R}$ defined by the entries $\tilde
R_{i,j}=R_{i,j}\left(\Av{\Psi_j}/\Av{\Psi_i}\right)^{1/2}$~\cite{Bicout1998,Hummer2005}. For
a process described by Eq.~\ref{Eq:MasterEqFP}, the likelihood of
observing a certain sequence
$\{Q^{(i_{\alpha})}(t_\alpha)\}_{\alpha=0}^N$ with $N$ transitions at
equidistant time intervals $\Delta t$ is:
\begin{equation}
\label{Eq:Likelihood}
 L=\prod_{\alpha=1}^N p(i_{\alpha}\vert i_{\alpha-1};t_{\alpha}-t_{\alpha-1})=\prod_{i,j=1}^M p(i\vert j;\Delta t)^{N_{ij}},
\end{equation}
where $N_{ij}$ is the total number of transitions from $j$ to $i$
observed along the trajectory and the time intervals
$t_{\alpha}-t_{\alpha-1}=\Delta t\;\forall\,\alpha$. Bayesian
inference (BI) can be used to determine the underlying free energy
$F(Q)$ and diffusivity profile $D(Q)$ from a stochastic
trajectory. Bayes' theorem states that for a given trajectory
($\equiv\text{data}$) the probability of certain parameters $\{F,D\}$
to be correct is:
\begin{eqnarray}
\label{Eq:Bayes}
 p(\{F,D\}\vert \text{data})&=&\frac{p(\text{data}\vert\{F,D\})\cdot p(\{F,D\})}{p(\text{data})} \nonumber \\
&\propto& L \cdot \underbrace{\prod_{i=1}^{M-1}\exp{\left(-\frac{(D_{i+1}-D_i)^2}{2\gamma^2}\right)}}_{\equiv p(\{F,D\})},
\end{eqnarray}
where $L$ is the likelihood of Eq.~\ref{Eq:Likelihood}; in our case the prior $p(\{F,D\})$ just depends on the diffusivity profile $D(Q)$, penalizing large deviations of the diffusivity at adjacent grid points.

\subsection{Optimization procedure}
A standard simulated annealing scheme is used to optimize the probability $p(\{F,D\}\vert \text{data})$ in Eq.~\ref{Eq:Bayes} by iterative variation of the $2M-1$ parameters of the system. The quantity to be minimized is the ``energy'' $E$ defined by: 
\begin{equation}
\label{Eq:Energy}
 E\equiv-\frac{\log(L)}{N}-\log(p(\{F,D\})).
\end{equation}
At each step the parameters $\{R_{i,i+1}\}_{i=1}^{M-1}$ and
$\{P_{i}\}_{i=1}^M$ are slightly perturbed giving rise to a new
configuration with energy $E^\text{new}$, which is always accepted for
$E^\text{new}\leq E$ and accepted with probability
$p^\tx{acc}=\exp{\left(-(E^\text{new}-E)/T\right)}$ for
$E^\text{new}>E$; the ``temperature'' $T$ of the system is
subsequently lowered until the optimized $F(Q)$ and $D(Q)$ are
reached. Several independent simulated annealing runs are performed;
variations in the results obtained in different runs allow drawing
conclusions on the quality of the estimate and the suitability of the
process for a FP type description.

\begin{figure}
       \includegraphics*{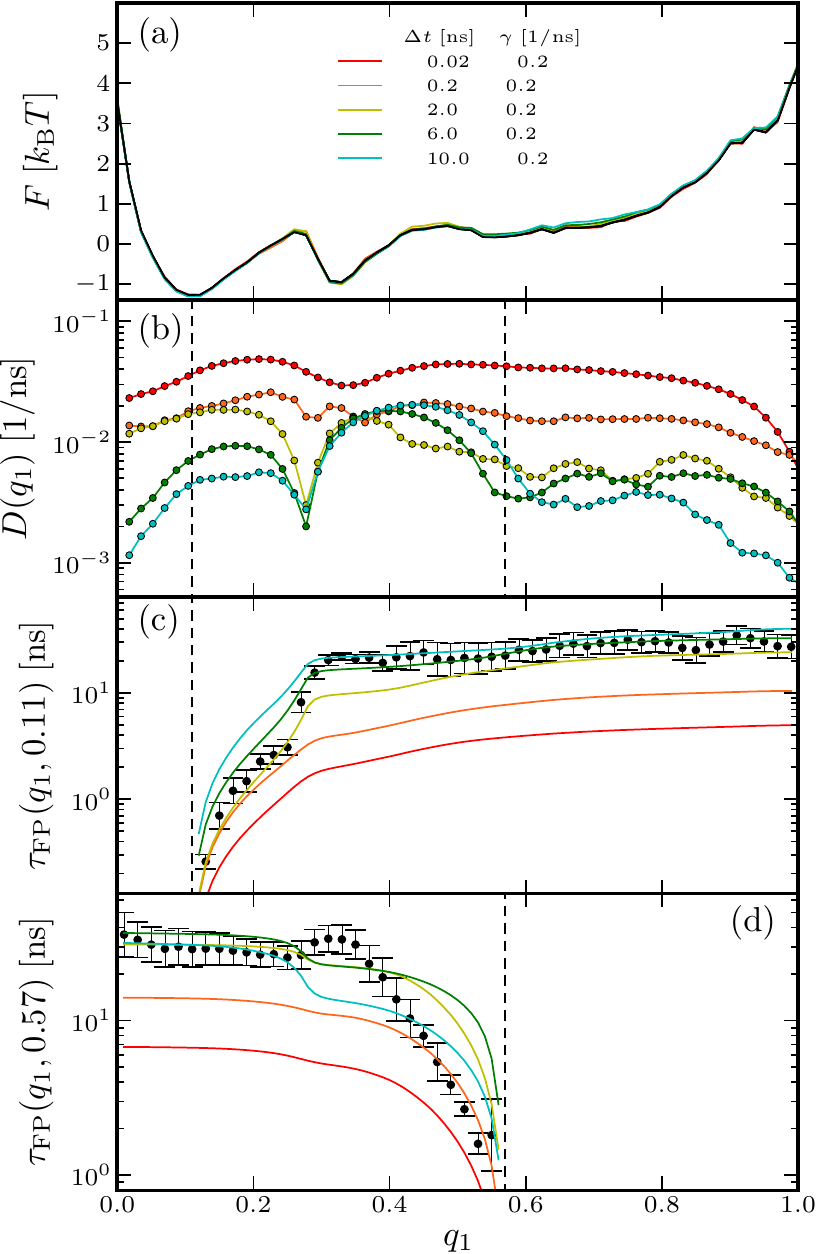}
      \caption{Free-energy $F$ (a), diffusivity $D$ (b) and MFP times
        $\tau_\tx{FP}(q_1,q_1^f)$ (panels (c) and (d)) for fixed value
        of $\gamma=0.2/\tx{ns}$ and different times intervals $\Delta
        t$ used in the optimization procedure.  The free-energy
        obtained from the equilibrium analysis of the trajectory is
        shown as a solid black curve in (a), the target states
        $q_1^f=0.11$ and $q_1^f=0.57$ are denoted as vertical dashed
        black lines, and the values of $\tau_\tx{FP}(q_1,q_1^f)$
        extracted directly from the simulation data as black circles
        in (c) and (d).}
      \label{Fig:Lagtime}
\end{figure}
\begin{figure}
      \includegraphics*{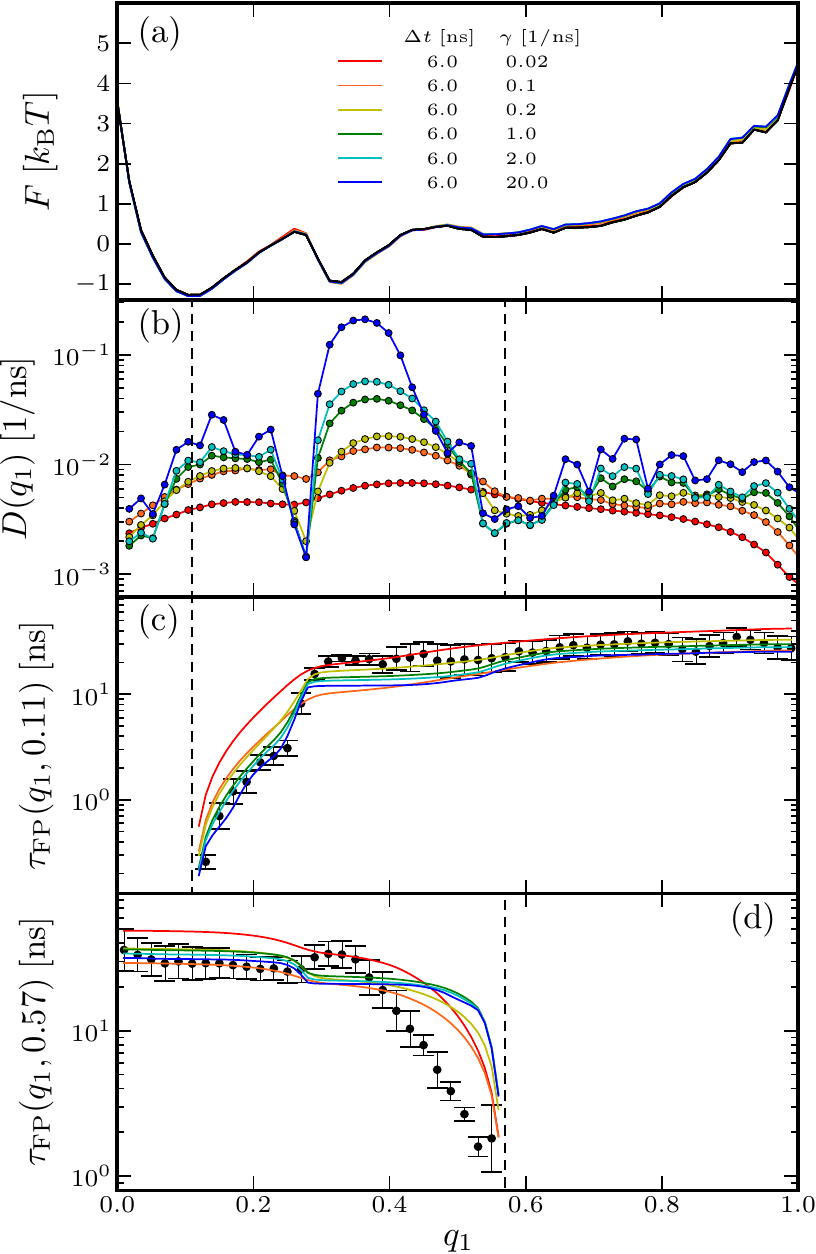}
      \caption{Same as Fig.~\ref{Fig:Lagtime}, but showing the
        influence of the smoothing parameter $\gamma$ on the
        diffusivity profile $D(q_1)$ for fixed time interval $\Delta
        t=6~\tx{ns}$.}
      \label{Fig:Gamma}
\end{figure}

\subsection{Dependence of $D(Q)$ on the time interval $\Delta t$ and the smoothing parameter $\gamma$}

The Bayesian optimization method is applied to the dynamics of the
reaction coordinate $Q\equiv q_1$. In Fig.~\ref{Fig:Lagtime} we
compare results obtained for different time intervals $\Delta t$; in
Fig.~\ref{Fig:Gamma} results for different values of the smoothing
parameter $\gamma$ weighting the prior in Eq.~\ref{Eq:Bayes} are
shown.

We show results for $60$ bins along the RC, and show average values of
$F(q_1)$ and $D(q_1)$ from $50$ independent optimization runs. Though
being a fit quantity, the free energy profile $F(q_1)$ does not
significantly differ from $\Av{\Psi_i}$ obtained from the equilibrium
analysis of the trajectory (black lines in the upper panel of the
figures). We note that the diffusivity profile $D(q_1)$ is strongly
sensitive on the time interval $\Delta t$: while almost identical
profiles like in the variance method analysis are obtained for $\Delta
t=20~\tx{ps}$, the diffusivity subsequently decreases for larger
$\Delta t$. The parameter $\gamma$ can compensate insufficient
sampling by externally requiring a smoothness of the diffusivity;
however, strong external constraints corresponding to low
$\gamma$-values tend to erase any structure in $D(q_1)$.

Reasonable choices of the parameters $\gamma$ and $\Delta t$ are not evident a priori --- to ensure that the long-time dynamics are correctly reproduced by the optimization result, we compute the position dependent MFP times $\tau_\text{FP}(q_1,q_1^f)$ for a folded state ($q_1^f=0.11$) and an unfolded one ($q_1^f=0.57$) for each of the optimized diffusivity profiles and compare these curves to the one directly extracted from the simulation data. This comparison shows that in our case $\gamma=0.2/\tx{ns}$ and $\Delta t=6~\tx{ns}$ are sensible values.

\end{document}